\documentstyle[12pt,epsfig]{article}                                                    

\newlength{\dinwidth}                                                    
\newlength{\dinmargin}                                                    
\def\lapproxeq{\lower .7ex\hbox{$\;\stackrel{\textstyle                                                    
<}{\sim}\;$}}                                                    
\def\gapproxeq{\lower .7ex\hbox{$\;\stackrel{\textstyle                                                    
>}{\sim}\;$}}                                                    
\def\be{\begin{equation}}                                                    
\def\ee{\end{equation}}                                                    
\def\bea{\begin{eqnarray}}                                                    
\def\eea{\end{eqnarray}}                                                    
\def\GeV{\rm GeV}
                                                    
\def\funp{{I\!\!P}}                    
                                                    
\begin{document}                                                    
\titlepage                                                    
\begin{flushright}                                                    
IPPP/07/66   \\
DCPT/07/132 \\                                                    
12 December 2007 \\                                                    
\end{flushright}                                                    
                                                    
\vspace*{2cm}                                                    
                                                    
\begin{center}                                                    
{\Large \bf Soft diffraction at the LHC: a partonic interpretation}                                                    
                                                    
\vspace*{1cm}                                                    
M.G. Ryskin$^{a,b}$, A.D. Martin$^a$ and V.A. Khoze$^a$ \\                                                    
                                                   
\vspace*{0.5cm}                                                    
$^a$ Institute for Particle Physics Phenomenology, University of Durham, Durham, DH1 3LE \\                                                   
$^b$ Petersburg Nuclear Physics Institute, Gatchina, St.~Petersburg, 188300, Russia            
\end{center}                                                    
                                                    
\vspace*{2cm}                                                    
                                                    
\begin{abstract}                                                    
We present a `new generation' model for high energy proton-proton `soft' interactions. It allows for a full set of multi-Pomeron vertices, as well as including multichannel eikonal scattering. It describes the behaviour of the proton-proton total, $\sigma_{\rm tot}$, and elastic $d\sigma_{\rm el}/dt$, cross sections together with those for low and high-mass proton dissociation. Although the model contains a comprehensive set of multi-Pomeron diagrams, it has a simple partonic interpretation.  Including the more complicated multi-Pomeron vertices reduces the absorptive effects as compared to the predictions in which only the triple-Pomeron vertex is considered. Tuning the model to describe the available `soft' data in the CERN ISR - Tevatron energy range, we predict the total, elastic, single- and double-diffractive dissociation cross sections at the LHC energy. An inescapable consequence of including multichannel eikonal and multi-Pomeron effects is that the total cross section is expected to be lower than before: indeed, we find $\sigma_{\rm tot}\simeq 90$ mb at the LHC energy. We also present differential forms of the cross sections. In addition we calculate soft diffractive central production.  
\end{abstract}                                          
     
%\newpage           
\section{Motivation}
It is essential to have a good model for the soft interactions of hadrons at high energies to, in particular, predict at the LHC, (i) the structure of underlying events, (ii) the value of the total cross section, $\sigma_{\rm tot}$ and the behaviour of the elastic cross section, $d\sigma_{\rm el}/dt$, and (iii) the probability of diffractive dissociation. Moreover, it is extremely important to understand the asymptotic behaviour of high energy interactions in a partonic framework. This will give the possibility of simulating the underlying events using a Monte Carlo generator based on a theoretically justified description of soft interactions. In addition, we need such a model to account for the effects of the underlying event in high energy hard interactions and to calculate the survival factors of rapidity gaps in exclusive and other hard diffractive processes. Finally it is crucial to have a good understanding of the spectra of leading nucleons, that is of the diffractive dissociation cross section, in order to describe extensive air showers and to interpret the highest energy cosmic rays.

\section{Overview of existing descriptions of soft interactions}
First we have the very simple Donnachie-Landshoff parametrization \cite{DL} of the high energy elastic amplitude, which is described in terms of simple poles in the complex angular momentum plane: namely the Pomeron and secondary Regge poles. The latter give a negligible contribution at Tevatron and LHC energies. However this parametrization says nothing about the distribution of secondary particles in the underlying event. Moreover, already at the LHC energy, $\sqrt{s}=14$ TeV, the amplitude $A(b,s)$ violates the black disc unitarity limit at small impact parameters $b \to 0$.         
       
We need to satisfy, at least, the two-particle $s$-channel unitarity relation, in order to respect the Froissart bound, and to describe the elastic cross section, which is about 20-25$\%$ of the total cross section in the Tevatron to LHC energy range. This leads to the eikonal form of the elastic amplitude, see (\ref{eq:elastamp}) in Section 3. To allow for the possibility of proton excitations we need to consider multichannel (say $n$-channel) eikonal models which include rescattering in the $i=1,...n$ diffractive eigenstates \cite{Good,kaid,tmr}. We will review the eikonal approach in the next Section. At present one- and two-channel eikonal models are used to predict the LHC cross sections \cite{uri, bh, KMRsoft}. A one-channel approach was used in \cite{bh,islam}, and two-channel eikonals were used in \cite{KMRsoft,uri,GLMnew}.

\begin{figure}
\begin{center}
\includegraphics[height=3cm]{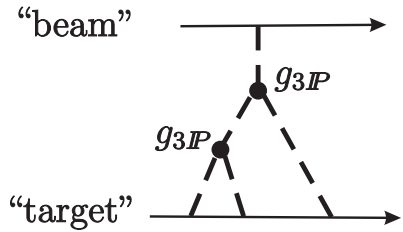}
\caption{A typical multi-Pomeron fan diagram contributing to the interaction between a small size beam particle and a large size target particle.}
\label{fig:fan}
\end{center}
\end{figure}

However, even a multichannel eikonal is unable to account for diffractive dissociation into high-mass states. These processes are usually described in terms of Regge theory with the help of the triple-Pomeron vertex. The problem is that, after we allow for the low values of the probability, $S^2$, that the rapidity gaps survive the eikonal rescattering, the value of the triple-Pomeron coupling, $g_{3\funp}$, needed to describe the data ($\sigma \sim S^2 g_{3\funp}$), becomes rather large; namely $g_{3\funp} \simeq g_N /3$ \cite{KMRj}, where $g_N$ is the nucleon-Pomeron coupling\footnote{Earlier estimates \cite{kkt,ff,kaid}, which do not account for the rescattering factor $S^2$, give $g_{3\funp} \simeq g_N /10$. However, as discussed in \cite{abram,KPT}, when we account for more complicated enhanced diagrams we need a larger value of the bare triple-Pomeron coupling.}. As a result we cannot neglect more complicated multi-Pomeron diagrams containing a large number of triple-Pomeron couplings. A specific set of such diagrams, known as ``fan'' diagrams, have been summed in Refs.~\cite{schw,glr,bk}. The selection of these fan diagrams is justified for proton-nuclei interactions or deep inelastic scattering.  That is when the size of the incoming object is much smaller than the size of the ``target''. In other words, we account for the multiple interactions with the large size ``target'', but not with the ``beam'' particle, see Fig.~\ref{fig:fan}. The summation of these diagrams leads to saturation, that is the amplitude $T \to$ constant at large $s$. The problem is that the fan diagram approximation can only be used to describe the beginning of the approach to saturation. Once we are near saturation it is not justified to neglect more complicated multi-Pomeron graphs\footnote{A more general, but still incomplete, set of multi-Pomeron graphs generated by $g_{3\funp}$ was considered in Ref. \cite{afs}, and more recently in Refs. \cite{ost,motyka}; note that in \cite{ost}, besides $g_{3\funp}$, more complicated multi-Pomeron vertices were considered}.

A good description of CERN ISR--Tevatron data on $\sigma_{\rm tot}, d\sigma_{\rm el}/dt$ was obtained in \cite{KMRsoft}, where the high-mass diffractive dissociation was described phenomenologically and added to the two-channel eikonal. That is, high-mass diffraction was not generated from the underlying theory, but was included in terms of the {\it leading} triple-Pomeron diagram, and then this contribution was added to the total proton opacity $\Omega (b)$.

The shortcomings of the existing approaches to describe soft interactions can be summarised as follows. First, it is not clear, a priori, how the results will change in going from a two- to a three- (or more) channel eikonal model. Furthermore, we do not know how the results depend on the size (that is, the form factor) of each diffractive eigenstate\footnote{In \cite{KMRsoft} both eigenstates are assumed to have the same size, and in the latest versions of the Tel-Aviv model \cite{GLMnew} another extreme is considered --- the size of the second component is zero.} $i$. Moreover, since $g_{3\funp}$ is not small, we cannot neglect the more complicated multi-Pomeron interactions\footnote{Here we seek a model for the high energy $pp$ interaction and therefore we neglect secondary Reggeon contributions, and include only the Pomeron.}.

Here we will consider a three-channel eikonal model. We will assume that the cross section of each eigenstate, $\sigma_i$ (that is, the coupling of the eigenstate to the Pomeron) is proportional to the square of the transverse size of the component, $R_i^2$. This form is motivated, {\it either} by leading-order QCD, where $\sigma \sim \alpha_S^2R^2$, {\it or} by the assumption that each eigenstate has an opacity of the same shape with the same value of $\Omega_i(b=0)$, such that the integrated coupling $\int d^2b~ \Omega_i \propto R_i^2$. Another possibility, which we will analyse, 
%is $\sigma_i \sim R_i$. Here the motivation is that the BFKL framework leads to an anomalous dimension 
%$\gamma \to 0.5$ of the amplitude at high energies \cite{BFKLrev}.
is to use the same form factor, that is the same size, for each component, as in \cite{KMRsoft}.

By allowing for a different size for each component, we have the possibility to introduce, and to calculate the different partonic composition of each diffractive eigenstate. For instance, for a smaller size component, the evolution will start at a larger scale $\mu$, and so there will be less gluons and other low $x$ partons radiated during the evolution.

\section{Eikonal model for diffractive scattering}
Let us delay the inclusion of high-mass diffractive dissociation for the moment in order to first introduce the eikonal description of soft proton-proton interactions. This will allow low-mass diffractive excitations to be included. We will then have to describe how to incorporate the important contributions made by high-mass single and double proton dissociation. The high-mass dissociations have a crucial effect on the predictions for the LHC.

Unitarity plays a pivotal role in diffractive processes. The total
cross section is intimately related to the elastic scattering
amplitude and the scattering into inelastic final states via
$s$-channel unitarity, $SS^\dag = I$, or
\be {\rm disc}\ T \equiv T-T^\dag = iT^\dag T \label{eq:discT}\ee
with $S=I+iT$. 

\subsection{Single-channel eikonal model}

First, we briefly recall the relevant features of the single-channel eikonal model. That is we focus on elastic
unitarity. Then ``disc $T$'' is simply the discontinuity of $T$
across the two-particle $s$-channel cut. At high energies we have a sizeable inelastic component. The
$s$-channel unitarity relation is diagonal in the impact
parameter, $b$, basis, and may be written
\be 2 {\rm Im}\,T_{\rm el}(s,b) = |T_{\rm el}(s,b)|^2 + G_{\rm
inel}(s,b) \label{eq:a1} \ee
with
\bea \sigma_{\rm tot} & = & 2\int d^2b\, {\rm Im}\,T_{\rm el}(s,b) \label{eq:ot} \\
\sigma_{\rm el} & = & \int d^2b\,|T_{\rm el}(s,b)|^2 \\
\sigma_{\rm inel} & = & \int d^2b\,\left[2{\rm Im}\,T_{\rm
el}(s,b) - |T_{\rm el}(s,b)|^2\right]. \eea
These equations are satisfied by
\bea {\rm Im}T_{\rm el}(s,b) & = & 1-{\rm e}^{-\Omega/2} \label{eq:elastamp}\\
\sigma_{\rm el}(s,b) & = & (1-{\rm e}^{-\Omega/2})^2, \label{eq:el}\\
\sigma_{\rm inel}(s,b) & = & 1- {\rm e}^{-\Omega}, \label{eq:inel} \eea
where $\Omega(s,b)\geq 0$ is called the
opacity (optical density) or eikonal\footnote{Sometimes $\Omega/2$
is called the eikonal; for simplicity we omit below the real part of $T_{\rm el}$. At high energies, the ratio ${\rm ReT}_{\rm el}/{\rm Im}T_{\rm el}$ is small, and can be evaluated via a dispersion relation.}. From (\ref{eq:inel}), we see that $ \exp(-\Omega(s,b))$ is the probability that
no inelastic scattering occurs.

The well known example of scattering by a black disc, with
${\rm Im}T_{\rm el}=1$ for $b<R$, gives $\sigma_{\rm el} = \sigma_{\rm inel}
=\pi R^2$ and $\sigma_{\rm tot} = 2\pi R^2$. In general, we see
that the absorption of the initial wave due to the existence of
many inelastic channels leads, via $s$-channel unitarity, to
diffractive dissociation.

\subsection{Inclusion of low-mass diffractive dissociation}
So much for elastic diffraction. Now we turn to inelastic
diffraction, which is a consequence of the {\em internal
structure} of hadrons. This is simplest to describe at high
energies, where the lifetime of the fluctuations of a fast hadron is
large, $\tau\sim E/m^2$, and during these time intervals the
corresponding Fock states can be considered as `frozen'. Each
hadronic constituent can undergo scattering and thus destroy the
coherence of the fluctuations. As a consequence, the outgoing
superposition of states will be different from the incident
particle, and will most likely contain multiparticle states, so we
will have {\em inelastic}, as well as elastic, diffraction.

To discuss inelastic diffraction, it is convenient to follow Good
and Walker~\cite{Good}, and to introduce states $\phi_k$ which
diagonalize the $T$ matrix. Such eigenstates only undergo elastic
scattering. Since there are no off-diagonal transitions,
\be \langle \phi_j|T|\phi_k\rangle = 0\qquad{\rm for}\ j\neq k, 
\ee
a state $k$ cannot diffractively dissociate into a state $j$. We
have noted that this is not, in general, true for hadronic states, which are not eigenstates of the $S$-matrix, that is of $T$. To account for the internal structure of the hadronic states, we have to enlarge the set of
intermediate states, from just the single elastic channel, and to
introduce a multichannel eikonal. We will consider such an example
below, but first let us express the cross section in terms of the
probability amplitudes $F_k$ of the hadronic process proceeding via the
various diffractive eigenstates\footnote{The exponent exp$(-\Omega_k)$ describes the probability that the diffractive eigenstate $\phi_k$ is not absorbed in the interaction. Later we will see that the rapidity gap survival factors, $S^2$, can be described in terms of such eikonal exponents.} $\phi_k$.

Let us denote the orthogonal matrix which diagonalizes ${\rm
Im}\,T$ by $a$, so that
\be \label{eq:b3} {\rm Im}\,T \; = \; aFa^T \quad\quad {\rm with}
\quad\quad \langle \phi_j |F| \phi_k \rangle \; = \; F_k \:
\delta_{jk}. \ee
Now consider the diffractive dissociation of an arbitrary incoming
state
\be \label{eq:b4} | j \rangle \; = \; \sum_k \: a_{jk} \: | \phi_k
\rangle. \ee
The elastic scattering amplitude for this state satisfies
\be \label{eq:b5} \langle j |{\rm Im}~T| j \rangle \; = \; \sum_k
\: |a_{jk}|^2 \: F_k \; = \; \langle F \rangle, \ee
where $F_k \equiv \langle \phi_k |F| \phi_k \rangle$ and where the
brackets of $\langle F \rangle$ mean that we take the average of
$F$ over the initial probability distribution of diffractive
eigenstates. After the diffractive scattering described by
$T_{fj}$, the final state $| f \rangle$ will, in general, be a
different superposition of eigenstates from that of $| j \rangle$,
which was shown in~(\ref{eq:b4}). At high energies we may neglect
the real parts of the diffractive amplitudes. Then, for cross
sections at a given impact parameter $b$, we have
\bea \label{eq:b6} \frac{d \sigma_{\rm tot}}{d^2 b} & = & 2 \:
{\rm Im} \langle j |T| j \rangle \; = \; 2 \: \sum_k
\: |a_{jk}|^2 \: F_k \; = \; 2 \langle F \rangle \nonumber \\
& & \nonumber \\
\frac{d \sigma_{\rm el}}{d^2 b} & = & \left | \langle j |T| j
\rangle \right |^2 \; = \; \left (
\sum_k \: |a_{jk}|^2 \: F_k \right )^2 \; = \; \langle F \rangle^2 \\
& & \nonumber \\
\frac{d \sigma_{\rm el \: + \: SD}}{d^2 b} & = & \sum_k \: \left |
\langle \phi_k |T| j \rangle \right |^2 \; = \; \sum_k \:
|a_{jk}|^2 \: F_k^2 \; = \; \langle F^2 \rangle. \nonumber \eea
It follows that the cross section for the single diffractive
dissociation of a proton,
\be \label{eq:b7} \frac{d \sigma_{\rm SD}}{d^2 b} \; = \; \langle
F^2 \rangle \: - \: \langle F \rangle^2, \ee
is given by the statistical dispersion in the absorption
probabilities of the diffractive eigenstates. Here the average is
taken over the components $k$ of the incoming proton which
dissociates. If the averages are taken over the components of both
of the incoming particles, then in (\ref{eq:b7}) we must introduce a second index on $F$, that is $F_{ik}$, and sum over $k$ and $i$. In this case the sum is the
cross section for single and double dissociation.

Note that if all the components $\phi_k$ of the incoming
diffractive state $| j \rangle$ were absorbed equally then the
diffracted superposition would be proportional to the incident one
and  the inelastic diffraction would be zero.  Thus if, at very
high energies, the amplitudes $F_k$ at small impact parameters are
equal to the black disk limit, $F_k = 1$, then diffractive
production will be equal to zero in this impact parameter domain
and so will only occur in the peripheral $b$ region. A similar
behaviour already takes place in $pp$ (and $p\bar{p}$)
interactions at Tevatron energies. Hence the impact parameter
structure of inelastic and elastic diffraction is drastically
different in the presence of strong $s$-channel unitarity effects. The elastic amplitude originates mainly from the centre of the disk (that is, from small $b$), while dissociation comes from the periphery. Hence it is important to pay special attention to the periphery of the proton, in impact parameter,        
$b$, space.  First, large values of $b$ are responsible for the small $t$ behaviour of the        
amplitude.  Second, the large $b$ region, where the optical density (or opacity), $\Omega        
(b)$, becomes small, gives the major contribution to the survival probability of rapidity     
gaps.

It is clear that if we allow for the possibility of diffractive excitation, then we will enlarge the absorptive effect caused by the eikonal. As a consequence the results obtained by a one- and two-channel eikonal fit will be substantially different. However, once we fix the dispersion, (\ref{eq:b7}), that is, once we fix the ratio\footnote{Note that in terms of a limited number of eigenstates $\phi_i$, we can consider only `low-mass' proton excitations, within a limited interval of $M^2$.} $\sigma_{\rm SD}$(low-mass)$/\sigma_{\rm el}$, it turns out that the inclusion of a third channel will not practically change the result. Of course, with a three-channel eikonal we have too many parameters. We will therefore consider an extreme case where each channel has the same weight; $a_i=1/\sqrt{3}$ with $i=1,2,3.$.  Otherwise the contribution of the eigenstate with the smallest weight will be less visible and the situation will be close to the 2-channel model.   Next, we fix the value of the dispersion, (\ref{eq:b7}),
\be
\langle \beta_i^2 \rangle~-~{\langle \beta_i \rangle}^2~=~(1+\gamma^2){\langle \beta_i \rangle}^2.
\label{eq:disp}
\ee
As a result the total probability of dissociation (in the limit of small opacity, $\Omega \ll 1$),
\be
\sigma_{\rm SD}^{{\rm low}M}/\sigma_{\rm el}=2\gamma^2,
\label{eq:fac2}
\ee
 will be the same in the two- and three-channel cases. Here $\beta_i$ is the $\phi_i$-Pomeron coupling, so the amplitude $F_i$ is proportional to $\beta_i$; the factor of 2 in (\ref{eq:fac2}) accounts for low-mass excitations of both the colliding beam particles.

Note that to describe the elastic scattering data in terms of a multichannel eikonal model, we have to increase the size of the intercept, $\alpha_{\funp}(0)$, and to reduce the slope $\alpha'$, of the bare Pomeron trajectory, in comparison with the naive Donnachie-Landshoff amplitude, in order to compensate for the absorptive effects of multichannel eikonal rescattering \cite{kaid,KPT,KMRcalabria}.

It was checked that the effects of the pion loop in the bare Pomeron trajectory (see \cite{KMRsoft}) can be mimicked by increasing the value of $\alpha'$. Thus, if we do not want $1\%$ accuracy, we can replace our previous two-channel description \cite{KMRsoft} of the data by a simpler two-channel fit, which is suitable for the calculation of the rapidity gap survival factors for the different hard diffractive processes and to evaluate the soft cross sections $\sigma_{\rm tot}$, $d\sigma_{\rm el}/dt$ and $\sigma_{\rm SD}^{{\rm low}M}$.

\subsection{Survival factors}   

An essential ingredient, in the prediction of the rate of a particular diffractive process, is the calculation of the suppression due to multi-Pomeron exchanges\footnote{This is especially important for searches for New Physics signals in processes with tagged forward protons, see, for example, \cite{KMRprosp,HKRSTW}.}. We assume that the rapidity gap is large enough that it will be populated by secondaries from soft rescattering;  that is, in an inelastic soft interaction, there is a negligible probability for a fluctuation with zero multiplicity within the large gap interval. Then we can write the survival factor of the gap with respect to the $i-k$ soft interaction at fixed impact parameter $b$ as exp$(-\Omega_{ik}(b))$. This survival factor should be averaged over the full set of diffractive eigenstates and over the impact parameter $b$. If we do not detect the outgoing protons the survival factor caused by eikonal rescattering is given by
\be 
{\overline {S^2}} = \frac{{\displaystyle \sum_{i,k}\int} d^2b \,
|a_{pi}|^2~|a_{p'k}|^2~|{\mathcal M}_{ik}|^2~\exp(-\Omega_{ik}(s,b))}{{\displaystyle
\sum_{i,k}\int} d^2b\,|a_{pi}|^2~|a_{p'k}|^2~|{\mathcal M}_{ik}|^2}\,, \label{eq:c3}
\ee
where $|a_{pi}|^2$ and $|a_{p'k}|^2$ are the probabilities of finding the partonic
diffractive eigenstates $i \equiv |\phi_i\rangle$ and $k \equiv |\phi_k\rangle$ in the two colliding proton states
$|p\rangle$ and $|p'\rangle$, respectively, see (\ref{eq:b4}).  $|{\mathcal M}_{ik}|^2$ is the
probability of producing the particular final system from the incoming eigenstates $i$ and $k$. If one, or both, of the outgoing protons is observed, then we have to average the amplitude and not the cross section. For example when both protons are tagged with $p_T=0$ we have
\be 
{\overline {S^2}} = \left |\frac{{\displaystyle \sum_{i,k}\int} d^2b \,
|a_{pi}|^2~|a_{p'k}|^2~{\mathcal M}_{ik}~\exp(-\Omega_{ik}(s,b)/2)}{{\displaystyle
\sum_{i,k}\int} d^2b\,|a_{pi}|^2~|a_{p'k}|^2~{\mathcal M}_{ik}} \right |^2 \,. \label{eq:c3pp}
\ee

The survival probability, $S^2$, can depend on the types of active partons $j$ and the value
of their momenta, $x_j$ and $\vec k_{\perp j}$.
In a multichannel eikonal model, with a few eigenchannels, we expect
the channel with the smallest cross section to contain mainly valence quarks with larger $x$, while the channels with larger cross sections will be due
to sea quarks and gluons, concentrated at smaller values of
$x$ \cite{KKMRprob}.

\section{Inclusion of high-mass diffraction}

\begin{figure}
\begin{center}
\includegraphics[height=3cm]{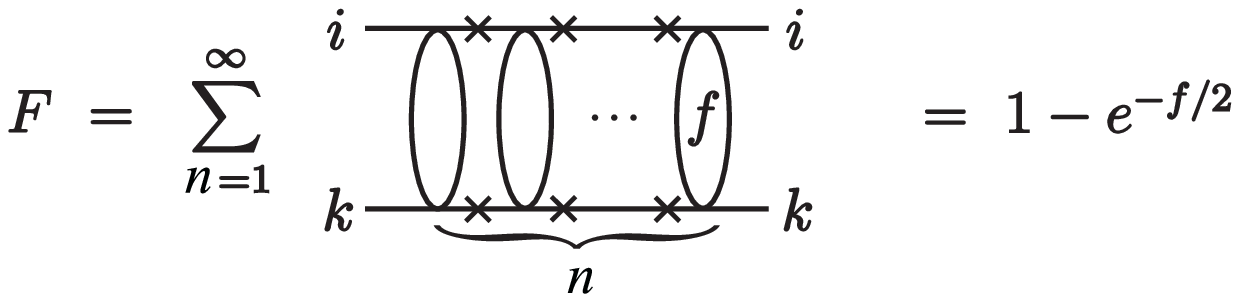}
\caption{The connection between the full amplitude $F_{ik}$ and the irreducible amplitude $f_{ik}$ for scattering between $i$ and $k$ diffractive eigenstates.}
\label{fig:fF}
\end{center}
\end{figure}

\begin{figure}
\begin{center}
\includegraphics[height=4cm]{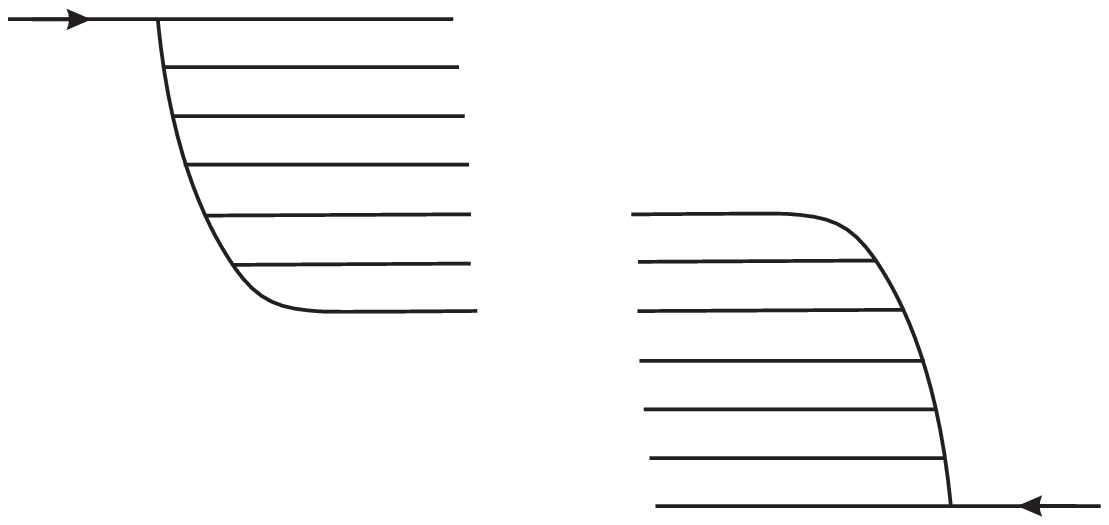}
\caption{A `soft' high energy interaction in which partons originating from the dissociations of the colliding protons overlap in rapidity. The overlap illustrates the impossibility of describing high-mass diffraction in terms of a pure eikonal (Good-Walker) formalism. Note that the master equations which describe the evolution of these `parton showers' are introduced in Section 4.2.}
\label{fig:olap}
\end{center}
\end{figure}

In the eikonal approximation the interaction of two diffractive eigenstates, $\phi_i$ and $\phi_k$ was described by the opacity $\Omega_{ik}$ corresponding to the two-particle $s$-channel irreducible amplitude. This $\Omega_{ik}$ generates the complete two-particle reducible amplitude $F_{ik}$,
\be
F_{ik}(s,b)~=~1-e^{-\Omega_{ik}/2},
\ee
see (\ref{eq:elastamp}). The connection between the irreducible amplitude $f_{ik}$ and the full amplitude $F_{ik}$ is illustrated in Fig.~\ref{fig:fF}. The defect of the approach, where the irreducible amplitude $\Omega_{ik} \equiv f_{ik}$ of the `$i-k$' interaction has been approximated by a single pole in the complex angular momentum plane, is that we are unable to separate the contribution coming from events with a rapidity gap from those of completely inelastic multiparticle processes. Thus we are unable to identify high-mass diffractive dissociation. Moreover, if we try to describe high-mass dissociation in the Good-Walker formalism we encounter the problem of double counting when the partons originating from the dissociation of the beam and `target' initial protons overlap in rapidities. These problems are well illustrated by the soft interaction sketched in Fig.~\ref{fig:olap}. In this Section we discuss how to treat high-mass diffractive dissociation. 

\subsection{The deficiency of the Schwimmer model}
\begin{figure}
\begin{center}
\includegraphics[height=2cm]{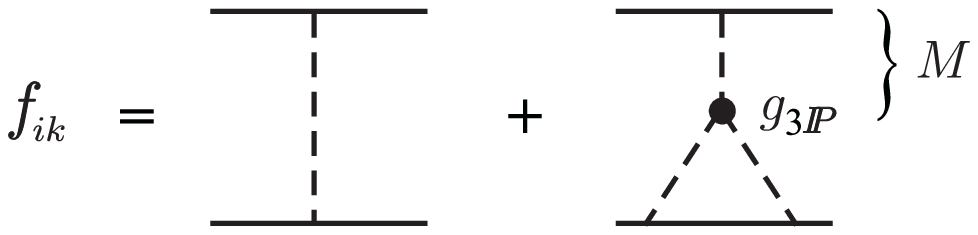}
\caption{The addition of the triple-Pomeron diagram which allows for high-mass, $M$, diffractive dissociation.}
\label{fig:add}
\end{center}
\end{figure}
\begin{figure}
\begin{center}
\includegraphics[height=3cm]{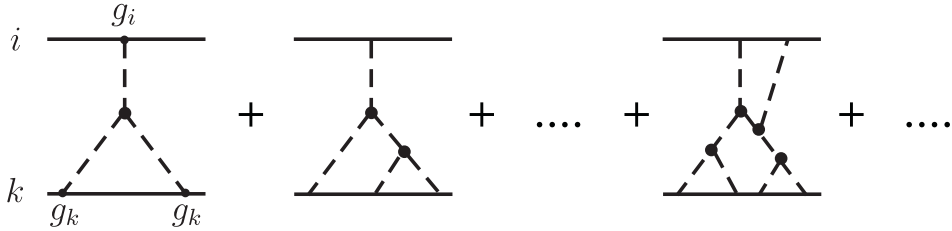}
\caption{The diagrams on the left-hand-side constitute the Schwimmer model. We also show a more complicated diagram.}
\label{fig:both}
\end{center}
\end{figure}
The usual way to include large mass $M$ dissociation is to include the triple-Pomeron diagram as in Fig.~\ref{fig:add}.  However, the triple-Pomeron coupling is not small, $g_{3\funp} \simeq g_N /3$ \cite{KMRj}, so to be consistent we have to sum up a series of more complicated graphs like those in Fig.~\ref{fig:both}. As mentioned above, if the size of one object ``$k$'' is much larger than that of the other, ``$i$'', then the major contribution comes from the sum of fan diagrams with multiple interactions on $k$. This is the so-called Schwimmer model \cite{schw}. In the Schwimmer model\footnote{For simplicity, we use the model in 1+1 space-time dimensions. We therefore use the notation ${\tilde g}_{3\funp}$ for the triple-Pomeron coupling in 1+1 dimensions in order to distinguish it from the real triple-Pomeron coupling $g_{3\funp}$.} the cross section becomes saturated at high energy, that is at high $Y$, \cite{schw,glm,BKKMR}
\be
\sigma_{\rm tot}~=~\frac{g_i g_k ~e^{\Delta Y}}{1+\epsilon (e^{\Delta Y}-1)}~~\to~~\sigma_{\rm tot}~=~\frac{g_i g_k}{\epsilon},
\label{eq:limtot}
\ee
where $Y={\rm ln}s$ is the available rapidity interval and $\Delta=\alpha_\funp (0)-1$; and where the value of $\epsilon={\tilde g}_{3\funp} g_k/\Delta$ fixes the maximum saturation density that can be reached in the Schwimmer model. At the same time, large mass diffractive dissociation is given by
\be
M^2\frac{d\sigma_{\rm SD}^{{\rm large}M}}{dM^2}~=~\frac{g_i g_k~\Delta~\epsilon\,~e^{\Delta (2 Y - {\rm ln}M^2)}}
 {[1+\epsilon\,(2e^{\Delta Y}-e^{\Delta (Y - {\rm ln}M^2)}-1)]^2} 
~~\to~~M^2\frac{d\sigma_{\rm SD}^{{\rm large}M}}{dM^2}~=~\frac{g_i g_k\Delta}{4\epsilon (M^2)^{\Delta}},
 \label{eq:limit}
\ee
where the final expression applies asymptotically, when $2\epsilon ~e^{\Delta Y} \gg 1$. By comparing (\ref{eq:limtot}) and (\ref{eq:limit}), we see that the absorptive effects are much stronger in diffractive dissociation than in the total cross section. Note the second power in the denominator in (\ref{eq:limit}) and, moreover the denominator contains 2exp($\Delta Y$) rather than exp($\Delta Y$). This reflects the well known fact that the screening of the diffractive dissociation amplitude is twice as strong as that for the elastic amplitude. Recall that eqs. (\ref{eq:limtot}) and (\ref{eq:limit}), and the discussion in the following subsections, 4.2 and 4.4, are in 1+1 dimensions.  Non-zero transverse momentum is introduced in Section 5.

\begin{figure}
\begin{center}
\includegraphics[height=10cm]{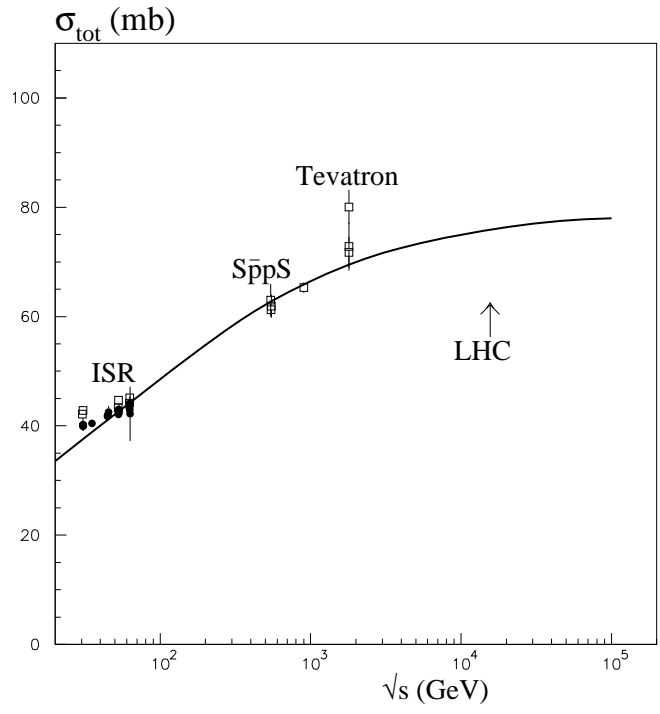} [t]
\caption{The description of the $pp$ total cross section in the Schwimmer model.}
\label{fig:sch}
\end{center}
\end{figure}

If, in the real world, we try to describe the total cross section in terms of the Schwimmer model we obtain a diffractive cross section, $d\sigma_{\rm SD}/dM^2$, which is an order of magnitude lower than that observed at the Tevatron. Indeed, as was discussed in Section 2, the triple-Pomeron vertex cannot be too small. Even neglecting the screening corrections, we need $g_{3\funp} \gapproxeq 0.1$. In particular, with $g_{3\funp}/g_N=0.1-0.2$ we find\footnote{The procedure that we use to include the transverse size of the proton, and other practical details, are described in Sections 5 and 6.} that the cross section $d\sigma_{\rm SD}/dx_L$ is about 5 times smaller than that observed by the CDF collaboration at the Tevatron. With a larger $g_{3\funp}$ the ``$\epsilon$'' term in the denominator of (\ref{eq:limit}) already becomes large for Tevatron energies; O(1) for $g_{3\funp} \gapproxeq 0.5$. Moreover, from the rightmost `limiting' expression in (\ref{eq:limit}), we see that $\sigma_{SD}$ is smaller for larger values of $g_{3\funp}$. For larger $g_{3\funp}=g_N/2$ we found that saturation is quickly obtained. However, as seen in Fig.~\ref{fig:sch}, it is still possible to describe the total and elastic cross section data.
Here, to compensate for the strong absorption, we need $\Delta=0.5$.  At the Tevatron energy\footnote{The Tevatron data that we use here were collected at an energy $\sqrt{s}=1.8$ TeV. For this reason we use a Tevatron energy of 1.8 TeV when we present our predictions.} this results in a diffractive cross section 
\be
(1-x_L)\frac{d\sigma_{\rm SD}}{dx_L dt}~=~0.2 ~{\rm mb}/{\rm GeV}^2,~~~~~~~ {\rm for} ~~~t=0,~~~x_L=0.99,
\ee
which is orders of magnitude less than that measured by CDF \cite{CDF,CDFsd}. 

Next, $d\sigma_{\rm SD}/dx_L$ decreases with energy as $s^{-\Delta}$, since for a fixed longitudinal beam momentum fraction, $x_L$, carried by the recoil proton\footnote{That is, for a fixed rapidity gap, $\Delta \eta={\rm ln}(1/(1-x_L))$.}, we have $M^2=(1-x_L)s$. Moreover, as discussed in \cite{KMRjhep}, the strong absorptive effect implied by the fan diagrams is not observed in the leading neutron spectra obtained at HERA. So the Schwimmer model is in contradiction with experiment.

\subsection{Including more complicated multi-Pomeron diagrams--zero transverse dimension}

\begin{figure}
\begin{center}
\includegraphics[height=3cm]{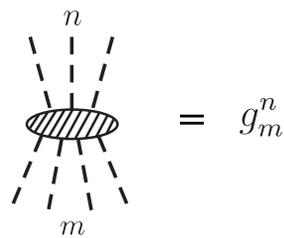}
\caption{A multi-Pomeron vertex}
\label{fig:gnm}
\end{center}
\end{figure}
For simplicity, we continue the discussion in 1+1 dimensions and wait until Section 5 to introduce non-zero transverse momentum. It is important to note that, besides the triple-Pomeron vertex, there are more complicated multi-Pomeron interactions, see Fig.~\ref{fig:gnm}. Indeed, it was proposed \cite{Cardy,KAT} to sum diagrams with all possible $(n,m)$ configurations at each vertex with coupling $g^n_m$. The exact values of the various $g^n_m$ vertices are, of course, unknown. In Ref.~\cite{Cardy} Cardy proposed a beautiful way to study the asymptotic high energy limit, by assuming the analyticity of $g^n_m$ in the Re $m$, Re $n ~\ge 0$ semi-plane. However, to describe the cross section at the energies available to experiment we need to know the explicit form of the vertices.  The simplest possibility is to assume an eikonal form
\be
g^n_m~ \propto ~g_N ~\lambda^{n+m-2}.
\ee
Such a programme was carried out in 1986 \cite{KPT}, and led to a rather reasonable prediction at the LHC energy. 

Here we will consider the partonic interpretation of an analogous approach. As mentioned above, for simplicity, we first neglect the transverse dimensions. In terms of the partonic evolution in rapidity space, the bare Pomeron pole contribution to the {\it elastic} amplitude $f$ is generated by the simple equation
\be
\frac{df(y)}{dy}~=~\Delta~ f(y).
\label{eq:T}
\ee
This equation can be regarded either as the equation for the single Pomeron amplitude, or as the equation for the parton density generated from either the beam $i$ or the target $k$. For definiteness, let us consider the partonic evolution from the target $k$. Then (\ref{eq:T}) gives
\be
f~=~g_k^2 ~e^{\Delta y},
\ee
where $\Delta$ corresponds to the intercept of the pole amplitude.  That is\footnote{Here, for fixed impact parameter, we use the same normalization of the amplitude as in (\ref{eq:ot}).}, $f \propto s^{\alpha -1}$, with $\alpha=1+\Delta$, so that $\sigma=g_k^2(s/s_0)^\Delta$, where $g_k$ is fixed by the initial condition, namely the probability of the interaction at $s=s_0$ --- in other words, by the parton density at $y=0$, the beginning of the evolution.

\begin{figure}
\begin{center}
\includegraphics[height=3cm]{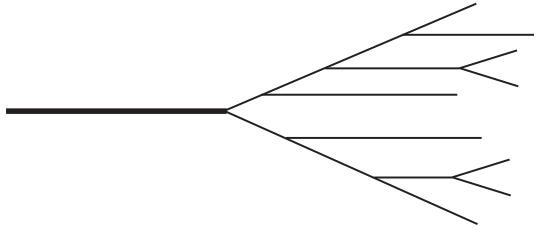}
\caption{A typical parton shower.}
\label{fig:shower}
\end{center}
\end{figure}
Here, at a low scale, it is impossible to say whether a parton is 
quark or gluon. These degrees of freedom are not well
defined in the `soft' regime\footnote{Nevertheless we hope that
at larger scales our partons will match smoothly with quark
and gluons.}.
However, in a soft high energy interaction, the parton may be
considered as a `small' elementary object which mediates
the process, in the spirit of the original parton model \cite{fey,Gribbook}. Thus, here, by the partonic picture
 we mean the evolution of the parton shower as
 shown in Fig. \ref{fig:shower}. We emphasize, that our approach goes beyond simply the
 probabilistic partonic interpretation of the amplitude, but, more
 important, it allows us to describe the actual evolution of
the parton shower through  master equations of the form of
(\ref{eq:T}) (or (\ref{eq:Teik}), (\ref{eq:Teik1}) and (\ref{eq:Teik2}) below).

\begin{figure}
\begin{center}
\includegraphics[height=3cm]{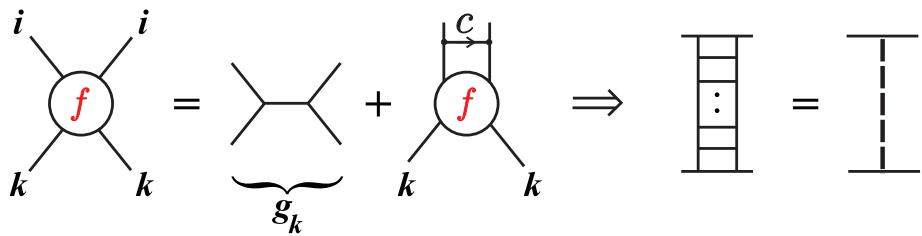}
\caption{The evolution of the elastic bare Pomeron amplitude}
\label{fig:dT}
\end{center}
\end{figure}
Indeed, according to (\ref{eq:T}), the `splitting function' $P$, which specifies the probability of the emission of an extra parton $c$ within the rapidity interval $dy$, is simply $P=\Delta$. When we iterate (\ref{eq:T}) we generate a ladder-type amplitude, which has exactly the structure of that considered in Ref.~\cite{amati}, but now in 1+1 dimensions. This is illustrated in Fig. \ref{fig:dT}, which produces an exchanged Pomeron (shown by the bold dashed line) with a ladder-type structure.

\begin{figure}
\begin{center}
\includegraphics[height=3cm]{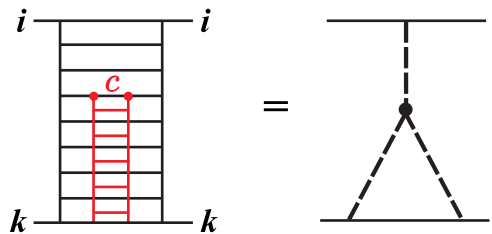}
\caption{The ladder structure of the triple-Pomeron amplitude.}
\label{fig:2lad}
\end{center}
\end{figure}

A multi-Pomeron contribution arises from the absorption of the intermediate $s$-channel partons $c$ during the evolution of $f$ in $y$. In particular the triple-Pomeron diagram in Fig.~\ref{fig:add} means that parton $c$ undergoes an extra rescattering with the target parton $k$, as shown in Fig.~\ref{fig:2lad}, \cite{Gribbook}. 
Allowing for many rescatterings, we have to sum over different numbers of ladders between partons $c$ and $k$. Assuming an eikonal form for the multi-Pomeron-proton vertex, it is natural to replace (\ref{eq:T}) by
\be
\frac{df}{dy}~=~f~\Delta~{\rm exp}(-\Omega/2), 
\label{eq:Teik}
\ee
where the `opacity' $\Omega \equiv \lambda f$ describes the transparency\footnote{Since the opacity depends on the type of incoming particle, it is natural to introduce a parameter $\lambda$ which reflects the difference of the opacity of the target felt by the intermediate parton and the opacity felt by the incoming eigenstate $i$.} of the target $k$. Since we are dealing with the elastic {\it amplitude} $f$ we use exp$(-\Omega/2)$ and not exp$(-\Omega)$, see eq. (\ref{eq:ot}). The coefficient $\lambda$ reflects the fact that parton $c$ may be different, and have a different absorptive cross section, from that of the eigenstate $i$. The value of this parameter $\lambda$ should be tuned to describe the data driven by the triple-Pomeron vertex. Since the triple-Pomeron vertex is relatively small, we expect $\lambda <1$.  Indeed, as we have already discussed, the HERA data on inelastic $J/\psi$ production (with target proton dissociation) indicate that \cite{KMRj} $g_{3\funp}/g_N \sim 0.3$. In the limit $\lambda \to 0$, we come back to the pole amplitude of (\ref{eq:T}).  For non-zero $\lambda$, it is straightforward to check that the asymptotic amplitude generated by (\ref{eq:Teik}) grows slowly as
\be
f~=~\frac{2}{\lambda}\left[{\rm ln}(y\Delta)~+~{\rm ln ~ln}y~+~....\right].
\label{eq:asymp}
\ee

This is very different to the Schwimmer model, which only accounts for the triple-Pomeron vertex. There we kept only the first two terms in the decomposition of the exponent in (\ref{eq:Teik}), so that the equation took the form
\be
\frac{df}{dy}~=~\Delta~f~-~\tilde{g}_{3\funp}~f^2. 
\label{eq:Tsch}
\ee
In that case the value of $f$ rapidly saturated, $f \to$ constant; see (\ref{eq:limtot}) with $\sigma_{\rm tot} \sim f$. The crucial difference is that in the perturbative (with respect to small $\tilde{g}_{3\funp}$) calculation, the right-hand-side of (\ref{eq:Tsch}) may even be {\it negative} for large $f$; while, after resummation of the multi-Pomeron exchanges, the right-hand-side of (\ref{eq:Teik}) is definitely {\it positive} \footnote{For the case of a {\it nuclear} target $k$, an equation analogous to (\ref{eq:Teik}) was discussed in \cite{lr}.}.

In terms of Regge diagrams, (\ref{eq:Teik}) sums up the system of fan diagrams in which any number $m$ of ``lower'' Pomerons couples to a fan vertex $g_m^1$, defined as in Fig.~\ref{fig:gnm}. In order to include the rescattering with the beam $i$ we replace (\ref{eq:Teik}) by
\be
\frac{df(y)}{dy}~=~f(y)~\Delta~e^{-(\Omega_k(y)+\Omega_i(y'))/2}.
\label{eq:Teik1}
\ee
The final term in the exponent is the opacity of the beam $i$, which depends on the rapidity interval $y'=Y-y$, with $Y={\rm ln}s$. The equation for the opacity $\Omega_i=\lambda f_i$ has the analogous form
\be
\frac{df(y')}{dy'}~=~f(y')~\Delta~e^{-(\Omega_i(y')+\Omega_k(y))/2}.
\label{eq:Teik2}
\ee
Actually, this is the same equation as (\ref{eq:Teik1}), but now evolving in the backward direction starting from the boundary condition $f(y'=0)=g_i$ at $y=Y$.

The system of equations (\ref{eq:Teik1}) and (\ref{eq:Teik2}), with boundary conditions $f(y=0)=g_k$ and $f(y'=0)=g_i$ may be solved by iteration \cite{ost,motyka}. Depending on the values of $g_i, ~g_k$ and $\Delta$, we usually need no more than $5-15$ iterations to reach an accuracy of $0.1\%$. That is the forward evolution of the amplitude $f(y)$ in the `background' field $f(y')$ gives the same result, to $0.1\%$ accuracy, as the backward evolution of $f(y')$ in the `background' field $f(y)$. Once (\ref{eq:Teik1}) and (\ref{eq:Teik2}) are solved, it is straightforward to simulate in a Monte Carlo the development of the parton shower (\ref{eq:Teik1}) in the known external field $\Omega_i$. 

\begin{figure}
\begin{center}
\includegraphics[height=3cm]{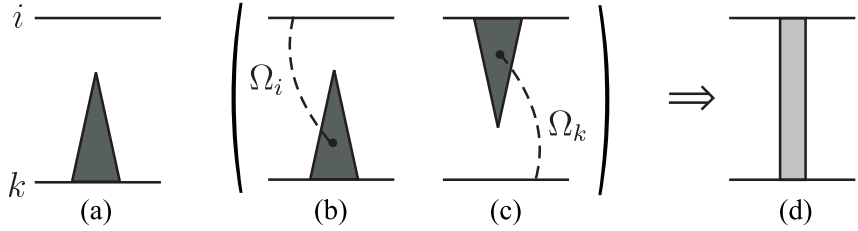}
\caption{Symbolic representation of the sum of multi-Pomeron diagrams generated by (a) equation (\ref{eq:Teik}), (b) equation (\ref{eq:Teik1}), (c) equation (\ref{eq:Teik2}), and (d) the system of equations (\ref{eq:Teik1}) and (\ref{eq:Teik2}).}
\label{fig:f7}
\end{center}
\end{figure}

At this stage, it is useful to have a pictorial illustration of the above equations. In terms of the diagrams of Reggeon Field Theory, (\ref{eq:Tsch}) produces the set of fan diagrams generated by the triple-Pomeron vertex only. The first two diagrams are shown in the left part of Fig.~\ref{fig:both}. On the other hand, (\ref{eq:Teik}) produces an analogous set of fan diagrams, but now the number of ``lower'' Pomerons in any vertex is arbitrary. Expanding the exponent on the right-hand-side of (\ref{eq:Teik}) leads to vertices $g^1_m$ with $m \ge 2$.  The factor $1/m!$ which comes from the expansion of the exponent accounts for the identity of the bare Pomerons\footnote{Strictly speaking, equations (\ref{eq:Teik}), (\ref{eq:Teik1}) and (\ref{eq:Teik2}) correspond to vertices with $g^n_m\propto nm\lambda^{n+m-2}$.}. The set of fan diagrams generated by (\ref{eq:Teik}) is shown symbolically in Fig.~\ref{fig:f7}(a) as a shaded triangle.  Now we turn to equations (\ref{eq:Teik1}) and (\ref{eq:Teik2}). The factor exp$(-\Omega_i(y')/2)$ in (\ref{eq:Teik1}), which describes the rescattering of the intermediate parton $c$ with the beam $i$, generates vertices like $g^n_1$, and leads to more complicated diagrams, one of which, for example, is shown by the last diagram in Fig.~\ref{fig:both}. If we expand both exponents exp$(-\Omega_k(y)/2)$ and exp$(-\Omega_i(y')/2)$, then we obtain diagrams with a full set of vertices $g^n_m$. In Fig.~\ref{fig:f7}(b), the effect of the absorptive factor exp$(-\Omega_i(y')/2)$ is shown by a dashed line. The opacity $\Omega_i$ is generated by (\ref{eq:Teik2}), which corresponds to Fig.~\ref{fig:f7}(c). Solving the system of equations (\ref{eq:Teik1}) and (\ref{eq:Teik2}) we obtain the full amplitude, which we show symbolically as the light-shaded rectangle in Fig.~\ref{fig:f7}(d).

\begin{figure}
\begin{center}
\includegraphics[height=3cm]{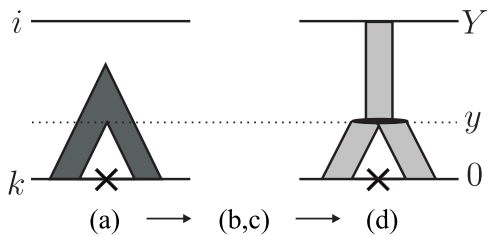}
\caption{The symbolic description of diagrams appropriate for diffractive dissociation of the eigenstate $i$ with a rapidity gap in some interval $(0,y)$.  The notation (a) $\to$ (b,c) $\to$ (d) refers to an analogous development to that indicated in Fig.~\ref{fig:f7}. The cross on the lower line indicates that eigenstate $k$ is on-shell.}
\label{fig:f8}
\end{center}
\end{figure} 
Note that the fan diagram of Fig.~\ref{fig:f7}(a) contains a contribution with a rapidity gap in some interval $(0,y)$, as shown in Fig.~\ref{fig:f8}(a). An analogous rapidity gap can be found in the whole amplitude Fig.~\ref{fig:f7}(d). We show this contribution in  Fig.~\ref{fig:f8}(d), where the light-shaded `rectangles' correspond to total amplitudes generated by (\ref{eq:Teik1}) and (\ref{eq:Teik2}). Note that in each `rectangle' in Fig.~\ref{fig:f8}(d) we account for the absorptive effect exp$(-\Omega /2)$ of the corresponding external field generated in the whole rapidity interval $(0,Y)$, and not just in the sub-intervals $(0,y)$ or $(y,Y)$ occupied by the particular `rectangle'.

\subsection{Comments on the final state}
Recall that in the evolution equations for amplitude, given in (\ref{eq:Teik}), (\ref{eq:Teik1}) and (\ref{eq:Teik2}), we
include the absorptive factor $\exp(-\Omega/2)$ and not
$\exp(-\Omega)$. That is we work with the forward amplitude ${\rm Im}
T=1-e^{-\Omega/2}$, which at each step of the evolution (in rapidity $y$)
includes all possible processes - both elastic and inelastic interactions with cross sections
$\sigma_{\rm el}=(1-e^{-\Omega/2})^2$ and 
$\sigma_{\rm inel}=1-e^{-\Omega}$; where $\sigma_{\rm tot}=2{\rm Im}
T=\sigma_{\rm el}+\sigma_{\rm inel}$, see eqs. (\ref{eq:elastamp})-(\ref{eq:inel}).

Strictly speaking using the AGK cutting rules \cite{agk}, together with
vertices $g^n_m\propto nm\lambda^{n+m-2}$, we would obtain
a parton-target elastic cross section
$\sigma_{\rm el}=(1-e^{-\Omega/2})(\Omega/2)e^{-\Omega/2}$, instead of form (\ref{eq:el}). On the other
hand the AGK rules were not proved for the general case of $g^n_m$ with
$n+m>3$. Here we prefer to adopt the partonic basis and to calculate the elastic
cross section following the usual relation (\ref{eq:el}).

As usual the inelastic processes includes both single-ladder
exchange, as well as multiple interactions with a larger density of
secondary partons. Analogous to the rescattering of a fast hadron in
a heavy nucleus, we assume that the probability, $w_N$, of events
with parton multiplicity $N$ times larger than that in a single
ladder, is given by
\be
 w_N=\frac{\Omega^N}{N!}e^{-\Omega}.
\ee
Unfortunately we cannot
use this probability $w_N$ literally to describe the
multiplicity distributions of secondary {\it hadrons}.

First, the distribution is affected by coherence effects.
For example, in quark-quark scattering, mediated by the exchange of
$N$ $t$-channel gluons, the colour flow between the quarks cannot exceed
the flow corresponding to octet (i.e. one-gluon) exchange. Thus, for
low $p_T$, when all $N$ gluons act coherently, we expect the
multiplicity of hadrons produced by hadronisation of the ($N$-gluon) colour tube to be the same as that for one-gluon
exchange, see also \cite{KLO}. Secondly, a non-negligible fraction of the final hadrons may be produced via the
fragmentation of minijets. These processes are beyond the `pure soft'
approach used in the present paper.
Therefore below we concentrate on the processes with rapidity
gaps.

\subsection{High-mass diffractive dissociation}
\begin{figure}
\begin{center}
\includegraphics[height=4cm]{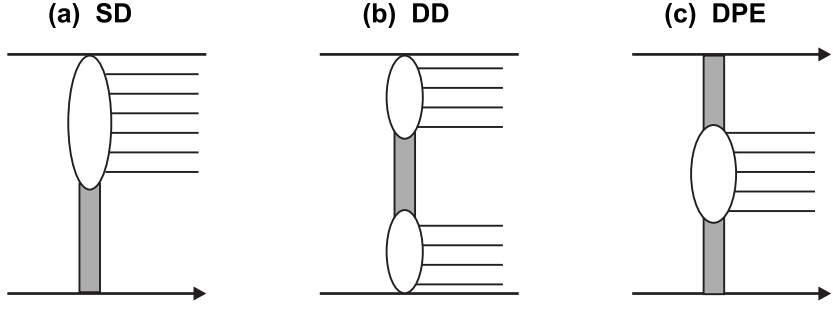}
\caption{The cross sections of these three high-mass diffractive processes are calculated in this paper. The processes are (a) the dissociation of a single proton, (b) double dissociation (DD) and (c) central production (DPE). The shaded rectangles represent the interactions within the rapidity gaps, and are the result of complicated multi-Pomeron effects, see Fig.~\ref{fig:f7}.}
\label{fig:dpe}
\end{center}
\end{figure}
We evaluate the cross sections for the three processes shown in Fig.~\ref{fig:dpe}, that is for the dissociation of a single proton (SD), for the dissociation of both incoming protons (DD) and for the central production of a system of mass $M$ separated from the outgoing forward protons by rapidity gaps (DPE). In naive simplified models the latter process is called double-Pomeron-exchange. In reality the rapidity gaps shown as shaded `rectangles' in Fig.~\ref{fig:dpe} are the result of complicated multi-Pomeron effects, see Fig.~\ref{fig:f7}. The results for the cross sections calculated in this paper correspond to rapidity gaps with $\Delta y>3$.

\begin{figure}
\begin{center}
\includegraphics[height=4cm]{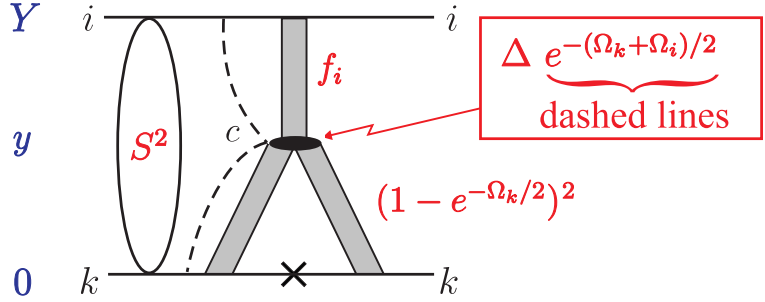}
\caption{The symbolic diagram for eq.(\ref{eq:dD0}) which describes single proton diffractive dissociation process $pp \to X+p$. The origin of the various factors in (\ref{eq:dD0}) are indicated.}
\label{fig:f9}
\end{center}
\end{figure} 
\subsection*{The cross section for single diffractive dissociation, $\sigma_{\rm SD}$}
We consider the diffractive dissociation of the beam particle $i$ into a system of large mass $M$ which occupies the rapidity interval from $Y$ to $y$. It arises from the elastic scattering of an intermediate parton $c$ on the target $k$, see Fig.~\ref{fig:f9}. This elastic cross section,
\be
\sigma ~ = ~ (1-e^{-\Omega/2})^2 
\ee
arises from the absorption $e^{-\Omega}$ of parton $c$ as a consequence of the solution of the unitarity equation (\ref{eq:a1}) for parton $c$, see (\ref{eq:el}). In this way we obtain a process with a rapidity gap in the interval $(0,y)$. The resulting cross section for single diffractive dissociation is
\be
\frac{d\sigma_{\rm SD}}{dy'}~=~(1-e^{-\Omega/2})^2~\Delta~ e^{-(\Omega_k~+~\Omega_i)/2}~f_i(y')~S^2
\label{eq:dD0}
\ee
with $y'=Y-y={\rm ln}M^2$. It is proportional to (i) the probability to find parton $c$ in the interval $dy'$, that is $\Delta~ e^{-(\Omega_k~+~\Omega_i)/2}$ of (\ref{eq:Teik2}); (ii) to the amplitude $f_i(y')$ of the parton $c$ - beam $i$ interaction; (iii) to the gap survival factor $S^2$, that is the probability $S^2 = e^{-f(Y)}$ to have no additional $i-k$ rescattering; and (iv) to the elastic $c-k$ cross section $(1-e^{-\Omega/2})^2$. The opacity $\Omega$ which drives this elastic amplitude is the opacity of the target, $\Omega_k(y)$.

\begin{figure}
\begin{center}
\includegraphics[height=4cm]{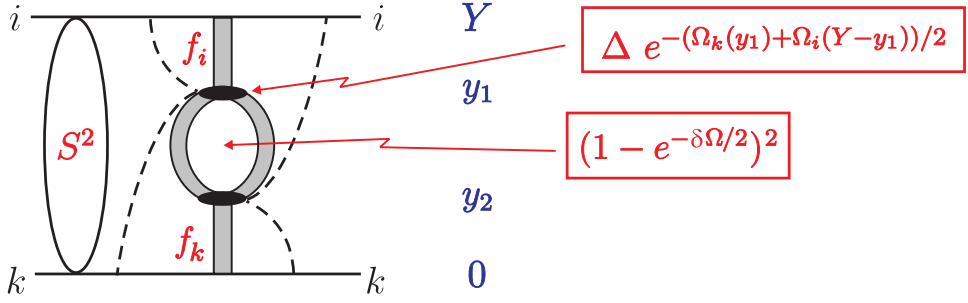}
\caption{The symbolic diagram for eq.(\ref{eq:DD1}) which describes the double diffractive dissociation process $pp \to X+X'$. The origin of the various factors in (\ref{eq:DD1}) are indicated.}
\label{fig:DD}
\end{center}
\end{figure} 
\subsection*{The cross section for double diffractive dissociation, $\sigma_{\rm DD}$}
Double diffractive dissociation (DD) can be calculated in an analogous way. Here we have to consider the `elastic' scattering of two intermediate partons, say parton $c$ at $y=y_1$ with $y'=Y-y_1$ and parton $d$ at $y=y_2<y_1$ with $y'=Y-y_2$. The corresponding opacity can be calculated as the difference between the {\it beam} opacities at $y=y_2$ and $y=y_1$
\be
\delta\Omega~=~\frac{\Omega_i(Y-y_2)-\Omega_i(Y-y_1)}{\Omega_i(Y-y_1)}~=~\frac{\Omega_i(Y-y_2)}{\Omega_i(Y-y_1)}~-~1.
\label{eq:DD3}
\ee
The denominator, $\Omega_i(Y-y_1)$, is the probability that an additional opacity $\delta\Omega$ is generated by the branch which contains the parton $c$. 
Alternatively the same quantity $\delta\Omega$ can be calculated from the {\it target} side
\be
\delta\Omega~=~\frac{\Omega_k(y_1)}{\Omega_k(y_2)}~-~1.
\ee
After the system of equations (\ref{eq:Teik1}) and (\ref{eq:Teik2}) is solved, both expressions give the same result, as may be checked numerically.

\begin{figure}
\begin{center}
\includegraphics[height=4cm]{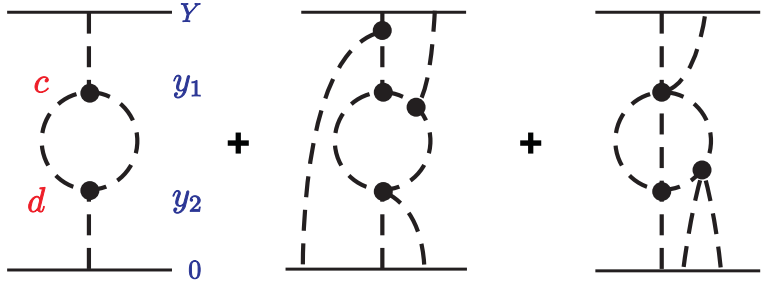}
\caption{Pomeron loop diagrams occurring in double diffractive dissociation.}
\label{fig:loops}
\end{center}
\end{figure}

The differential cross section for double dissociation is
\be
\frac{d^2\sigma_{\rm DD}}{dy_1dy_2}~=~(1-e^{-\delta\Omega/2})^2~\Delta e^{-(\Omega_k(y_2)+\Omega_i(Y-y_2))/2}~\Delta e^{-(\Omega_k(y_1)+\Omega_i(Y-y_1))/2}~f_i(Y-y_1)~f_k(y_2)~S^2.
\label{eq:DD1}
\ee
The amplitude $(1-e^{-\delta\Omega/2})$ describes the exchange of Pomerons between partons $c$ and $d$. In this way we allow for the Pomeron loop diagrams like those shown in Fig.~\ref{fig:loops}. The factor $1/n!$ which arises in the expansion of the exponential $e^{-\delta\Omega/2}$ accounts for the identity of the Pomerons arising in elastic $c-d$ scattering.

Double diffractive events, with a rapidity gap in the interval $(y_2,y_1)$, may also come from two simultaneous single dissociation interactions: one with a gap in the interval $(0,y_1)$ and the other with a gap in the interval $(y_2,Y)$. This contribution to the DD cross section is given by
\be
\frac{d^2\sigma_{\rm DD}^{\rm (SD*SD)}}{dy_1dy_2}~=~I_i(y_1)~I_k(y_2)~S^2,
\label{eq:DD2}
\ee
where
\be
I_i(y)~=~f_i(y')~(1-e^{\Omega_k(y)/2})^2~\Delta e^{-(\Omega_k+\Omega_i)/2} 
\ee
\be
I_k(y)~=~f_k(y)~(1-e^{\Omega_i(y')/2})^2~\Delta e^{-(\Omega_k+\Omega_i)/2},
\ee
are the probabilities of single dissociation (before the absorption), see (\ref{eq:dD0}); and $y'=Y-y$.

\subsection*{The cross section for central DPE production, $\sigma_{\rm DPE}$}
We can also calculate the cross section for the central DPE production of the process shown in Fig.~\ref{fig:dpe}(c). The central system has a mass given by $M^2~=~\xi_1~\xi_2~s$, where $\xi_i=(1-x_{iL})$ is the energy loss of the incoming protons. In comparison with the cross section of double dissociation of (\ref{eq:DD1}), where we consider the elastic scattering of the pair of intermediate partons $c$ and $d$, with rapidities $y_1$ and $y_2$, now we have the inelastic $c-d$ interaction accompanied by the elastic scattering of parton $d$ on the `target' $k$ and parton $c$ on the beam $i$. Thus the cross section is given by
\be
\frac{d^2\sigma_{\rm DPE}}{dy_1dy_2}~=~E_k~E_i~S^2~\frac{f_k(y_1)}{f_k(y_2)},
\label{eq:dpe}
\ee
with
\be
E_k~=~(1-e^{-\Omega_k(y_2)/2})^2~\Delta ~e^{-(\Omega_k(y_2)+\Omega_i(Y-y_2))/2}
\label{eq:40}
\ee
\be
E_i~=~(1-e^{-\Omega_i(Y-y_1)/2})^2~\Delta ~e^{-(\Omega_k(y_1)+\Omega_i(Y-y_1))/2},
\ee
where the $E_i$ and $E_k$ are the probabilities that the partons $c$ and $d$ participate in elastic interactions (multiplied by the probabilities to find these partons in the rapidity intervals $dy_1$ and $dy_2$ respectively). The rapidities are $Y-y_1=-\ln\xi_1$ and $y_2=-\ln\xi_2$. The last factor in (\ref{eq:dpe}) describes the probability that the $c-d$ interaction produces the central system of mass $M$;
in other words the factor $(1-e^{\Omega_k(y_2)/2})^2$ in (\ref{eq:40}),
which corresponds to elastic scattering, replaces the factor
$f_k(y_2)$ in the whole amplitude.

\subsection*{Asymptotic behaviour of diffractive dissociation}
At very high energies, $Y = {\rm ln} ~s\gg 1$, the whole irreducible amplitude given by (\ref{eq:Teik1}) and (\ref{eq:Teik2}) increases slowly
\be
f ~\simeq ~\frac{1}{\lambda}~{\rm ln}(Y\Delta),
\ee
see (\ref{eq:asymp}). On the other hand, the cross sections for single and double diffractive dissociation,
\be
\sigma_{\rm SD}~\sim~\frac{1}{\lambda}~(\Delta Y)^{-1/\lambda},~~~~~~~
\sigma_{\rm DD}~\sim~\frac{1}{\lambda^2}~(\Delta Y)^{-1/\lambda},
\ee
decrease as $(\Delta Y)^{-1/\lambda}$, due to the behaviour of the gap survival factor $S^2~=~e^{-f}~\sim~(\Delta Y)^{-1/\lambda}$. In the product $\Delta e^{-(\Omega_i+\Omega_k)/2}~f_i~\sim~1/\lambda Y$, the factor $Y$ in the denominator is compensated by the size of the available rapidity interval, $\int dy_1~\sim~Y$.

\section{Accounting for the size of the colliding hadrons}

Coming to the real 4-dimensional world, we may exploit the fact that the slope of the bare Pomeron trajectory needed to describe the elastic $pp$ data ($\alpha' \lapproxeq 0.1 ~\GeV^{-2}$) is very small in comparison with the slope of the proton form factor, $B_0 \simeq 2.5~\GeV^{-2}$. That is, the size of the Pomeron is much less than the size of the proton. Thus we may use the `` heavy Pomeron approximation'' \cite{gribovh}, solving the system of equations (\ref{eq:Teik1}) and (\ref{eq:Teik2}) written in zero transverse dimensions, but with initial conditions $g_k(b_k)$ and $g_i(b_i)$, which depend on the position of an effective heavy Pomeron with respect to the centre of the beam and target protons in impact parameter space. 

The $pp$ impact parameter is ${\vec b}={\vec b}_k-{\vec b}_i$, where here, and below, the vectors lie in the 2-dimensional transverse plane.  In $b$-space the input conditions for $f_i$ are given by the usual Fourier transform of the proton-Pomeron vertex $\beta_i$
\be
g_i({\vec b}_i)~=~\int \frac{d^2 q}{(2\pi)^2} ~\beta_i({\vec q}^{~2})~{\rm exp}(i{\vec q} \cdot {\vec b}_i),
\label{eq:ft}
\ee
and a corresponding identical equation for $g_k$. The whole irreducible amplitude $\hat{f}_{ik}$, which should be used to calculate the multichannel eikonal $\Omega_{ik}$ of Section 3.2, is given by
\be
\hat {f}_{ik}(Y,{\vec b})~=~\int d^2b_k d^2b_i~f_{ik}(Y;{\vec b}_i,{\vec b}_k)~\delta({\vec b}_k-{\vec b}_i-{\vec b})
\label{eq:A40}
\ee
with
\be
f_{ik}(Y,{\vec b}_i,{\vec b}_k)~~=~~f_k(y=Y,{\vec b}_k,{\vec b}_i)~g_i({\vec b}_i)~~=~~f_i(y'=Y,{\vec b}_i,{\vec b}_k)~g_k({\vec b}_k).
\ee
We keep both the arguments ${\vec b}_k$ and ${\vec b}_i$ since, at each impact parameter ${\vec b}$, we need to know the parton densities in both the beam and target protons. Indeed, the parton density $f_k(y)$ obtained from the evolution (\ref{eq:Teik1}) of the target partons depends not only on the impact parameter inside the target proton, ${\vec b}_k$, but also that, ${\vec b}_i$, for the beam proton through the external `background' field $\Omega_i (y')$ in (\ref{eq:Teik1}).  In terms of $\hat {f}_{ik}(Y,{\vec b})$, the total cross section takes the form
\be
\sigma_{\rm tot}~=~2\int d^2b~(1-e^{-\hat {f}_{ik}(Y,{\vec b})/2}),
\ee
see (\ref{eq:ot}) and (\ref{eq:elastamp}).  This completes our discussion of the elastic amplitude $f_{ik}$.  

To calculate the high-mass contribution to $\sigma_{\rm SD}$ we have just to replace $f_{ik}$ in (\ref{eq:A40}) by expression (\ref{eq:dD0}) for the `local' (that is fixed $b$) contribution to $\sigma_{\rm SD}$.
The cross section takes the form
\be
\frac{d\sigma_{\rm SD}}{dy'}~=~\int f_i(y')~(1-e^{-\Omega_k/2})^2~\Delta~ e^{-(\Omega_k~+~\Omega_i)/2}~S^2({\vec b}) ~d^2b_k d^2b_i
\label{eq:dD}
\ee
with $y'=Y-y={\rm ln}M^2$. In terms of the (experimentally measured) longitudinal momentum fraction $x_L$ of the leading proton, we have $dy'=dx_L/(1-x_L)$. In (\ref{eq:dD}), the eikonal gap survival factor $S^2$ depends on the $pp$ impact factor ${\vec b}={\vec b}_k-{\vec b}_i$, and the dependence of $f_i,~f_k$ on the $b$'s is implicit. Note that now the factor $S^2$ accounts for the possibility of rescattering at any point in the $b$-plane. That is
\be
S^2(b)~=~{\rm exp}(-\hat{f}_{ik}(Y,b)).
\ee 

The cross section of double diffractive dissociation is given by an analogous expression to (\ref{eq:dD}), but in which now the integrals have integrands given by (\ref{eq:DD1}) and (\ref{eq:DD2}). However, in the $\sigma_{\rm DD}^{\rm (SD*SD)}$ contribution of (\ref{eq:DD2}) we have to account for the possibility that the two {\it simultaneous} `single dissociations' $I_i$ and $I_k$ may take place at different impact parameters. That is we add the contribution
\be
\frac{d^2\sigma_{\rm DD}^{\rm (SD*SD)}}{dy_1dy_2}~=~\int d^2b~S^2(\vec {b})~\hat {I}_i(y_1,{\vec b})~\hat {I}_k(y_2,{\vec b}),
\label{eq:DD2DD}
\ee
where, in analogy to (\ref{eq:A40}),
\be
\hat {I}_j(y,{\vec b})~=~\int d^2b_k d^2b_i~I_j(y;{\vec b}_i,{\vec b}_k)~\delta({\vec b}_k-{\vec b}_i-{\vec b})
\label{eq:A40DD}
\ee
with $j=i,k$. The DPE cross section for central production is of the same form as (\ref{eq:dD}) but with (\ref{eq:dpe}) as the integrand.

We emphasize that everywhere in this section we are not dealing with the original proton beam and target, but rather with the Good-Walker diffractive eigenstates $i,k \equiv \phi_i,\phi_k$. Note that in this way we account for any low-mass intermediate states, that is we include all processes of the type $pp \to X+p,~pp \to X+N^*,...$. Finally we take the sum over the different eigenstates as described in Section 3. 

In terms of Reggeon Field Theory, the exponential factors exp$(-\Omega/2)$ create a very complicated system of multi-Pomeron exchange diagrams. However, as was mentioned in Section 3.1, actually the physical interpretation of such factors is simply the probability that {\it no} inelastic scattering occurs. Therefore, in terms of the parton approach, the equations generate only simple ladder diagrams, like Fig. \ref{fig:dT} or Fig. \ref{fig:2lad}, modified so that the probability (or splitting function) to produce a new parton (like $c$) is reduced by the `survival probability' of the parton in the background fields.

Now let us calculate the double differential diffractive cross section, $d^2\sigma_{SD}/dy'dq^2_T$, for $pp \to X+p$, where $q_T$ is the transverse momentum of the recoil proton. To do this we have to perform the Fourier transform of the $pp \to X+p$ amplitude with respect to the impact parameter $b_k$, that is
\be
\frac{d^2\sigma_{\rm SD}}{dy'dq^2_T}~=~\int \left|\int \sqrt {f_i(y')~(1-e^{-\Omega_k/2})^2~\Delta~ e^{-(\Omega_k+\Omega_i)/2}~S^2({\vec b})}~e^{i{\vec q}_T \cdot {\vec b}_k} ~d^2b_k \right|^2 \frac{d^2b_i}{4\pi}.
\label{eq:d2D}
\ee

\section{Practical details of the model}

Before we present the numerical results, it is convenient to list further details of the model.
\begin{itemize}
\item The low-mass excitations are included in the Good-Walker formalism (see Section 3). Therefore, to avoid double counting they should be omitted from the irreducible amplitude $f_{ik}$. Hence, we introduce a threshold $y_0$, and start (finish) the rapidity evolution of equations (\ref{eq:Teik1}) and (\ref{eq:Teik2}) at $y=y_0~(y=Y-y_0)$. We choose $y_0=2.3$, a value which is often used in Regge Field Theoretic calculations, and which effectively accounts for the next-to-leading log corrections to the BFKL equation, see \cite{KMRenhanced,KMRjhep}. Such a cut means that the low-mass dissociations with $M \lapproxeq 2.5$ GeV are accounted for within the Good-Walker eikonal formalism, while larger masses are described in terms of multi-Pomeron diagrams.
\item We consider two versions of the proton wave function, or, rather, of the $\phi_i$-Pomeron coupling $\beta_i(t)$:

\be
{\rm (A)}~~~~ \beta_i(t)~=~\beta_i(0)~V(t)~~~~~~~~~~~~~~~~
\label{eq:MA}
\ee
\be
{\rm (B)}~~~~ \beta_i(t)~=~\beta_i(0)~V(t\beta_i(0)/\langle \beta (0) \rangle)
\label{eq:MB}
\ee
where in the first version, each component $i$ has the same form factor, whereas in the second version the form factor of a component with a large cross section (large $\beta_i(0)$) is steeper than that for a component with a smaller cross section. In case (A) each component has the same transverse size and a larger cross section arises from a larger parton density. In version (B) the maximum parton density (which occurs at $b$=0) is the same for all components, and a larger coupling is caused by a larger transverse size.
\item The form factor of the $\phi_i$-Pomeron vertex is taken to have the form
\be
V(t)~= ~\frac{e^{at}}{(1-t/a_1)^2}.
\label{eq:V}
\ee
We take $a \simeq 0.1~\GeV^{-2}$ in the exponent in the numerator. It is introduced solely to provide better numerical convergence of the Fourier transform (\ref{eq:ft}).
\item The dispersion $\gamma^2$ of the coupling, (\ref{eq:disp}), is fixed by the experimental data on low-mass diffractive excitation. The analysis of the existing fixed target data require $\sigma_{\rm SD}^{{\rm low}M}/\sigma_{\rm el} \simeq 0.3$ \cite{kaid,KAID}. This value is consistent with the CERN ISR measurements of the excitations into particular channels ($N\pi, N\pi\pi, \Lambda K,...$) with $M<2.5$ GeV \cite{reson}, and corresponds to\footnote{Here, and in what follows, the value of $\sigma_{\rm SD}$ accounts for dissociation of both colliding particles.}
\be
\sigma_{\rm SD}^{{\rm low}M} \simeq 2~{\rm mb~~~~~~~~at}~\sqrt{s}=31~{\rm GeV}. 
\label{eq:2mb}
\ee
On the other hand in Ref.~\cite{ISR} the cross section extracted from measurements in the region $M^2/s<0.01$ was $\sigma_{\rm SD}^{{\rm low}M} \simeq 4$ mb for $s \simeq 1000~ \GeV^2$. However in this experiment the momentum resolution $\Delta M^2$ was comparable to the whole size of the $x_L$ interval, $\Delta x_L =0.01$. Thus the quoted value, $\sigma_{\rm SD}^{{\rm low}M} \simeq 4$ mb, includes some contribution from masses larger than 2.5 GeV and, more important, may also contain some admixture of elastic events. In order to study the possible effect of a larger cross section than (\ref{eq:2mb}), we repeat the entire analysis with $\sigma_{\rm SD}^{{\rm low}M} \simeq 3$ mb. (A value just below 3 mb was also measured in \cite{Albrow}.) Thus we will consider four versions of the model, which we denote by B2, B3, A2 and A3, where the letter A or B corresponds to the choice of couplings given by (\ref{eq:MA}) or (\ref{eq:MB}), and the number 2 or 3 refers to $\sigma_{\rm SD}^{{\rm low}M}$ in mb for $s \simeq 1000 \GeV^2$.
\item For both the $n=2$ and $n=3$ channel eikonals we take a simplified and extreme decomposition of the proton wave function
\be
|p \rangle~=~\frac{1}{\sqrt n}\sum^n_{i=1} |\phi_i \rangle.
\ee
\item The value of $\lambda$ was defined in Section 4.2 ($g^n_m \propto \lambda^{n+m-2}$ which leads to $\Omega\equiv \lambda f$). It controls the probability of high-mass diffractive dissociation. $\lambda$ was tuned to be consistent with the measured CDF data \cite{CDF,CDFsd} for $d\sigma_{\rm SD}/dx_Ldt$.
\item From the theoretical point of view the ``heavy Pomeron approximation'' is not totally satisfactory. To satisfy $t$-channel unitarity we need a non-zero slope of the Pomeron trajectory. At least, the two-pion loop contribution to Pomeron exchange should be included to satisfy unitarity at the nearest $t=4m^2_{\pi}$ branch point. However this contribution turns out to be small. We therefore set the slope $\alpha'=0$, which provides reasonably agreement with the data in the region of interest. Since we take $\alpha'=0$, we expect to underestimate the elastic slope, $B_{\rm el}$ at the LHC by up to 2$\%$. Finally, the intercept of the bare Pomeron, $\alpha_\funp (0)$, is left as a free parameter.
\end{itemize}

\section{Results of the model}

\begin{figure}
\begin{center}
\includegraphics[height=17cm]{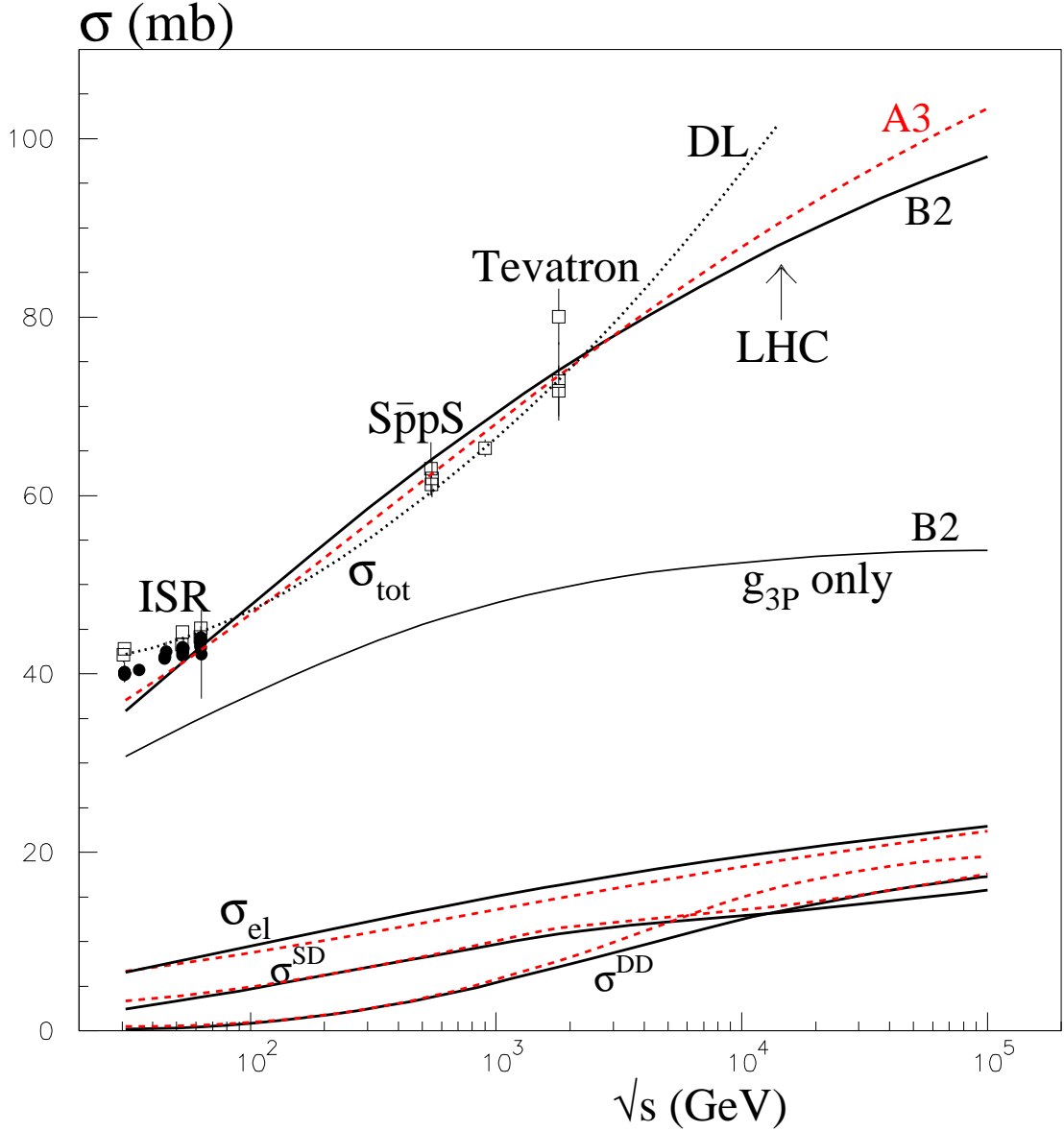}
\caption{The description of the total and diffractive dissociation $pp$ cross sections, obtained with a three-channel eikonal, for model (B2), shown by the continuous curves, and model (A3) shown by the dashed curves. The references to the data are as in \cite{KMRsoft}. Also shown by the curve denoted DL is the naive expectation for $\sigma_{\rm tot}$ obtained with a simple effective Pomeron pole (and secondary Regge contributions) \cite{DL}. The curve marked ``$g_{3\funp}$ only'' is to demonstrate the importance of higher multi-Pomeron contributions, see the text.}
\label{fig:r1}
\end{center}
\end{figure}

\begin{figure}
\begin{center}
\includegraphics[height=17cm]{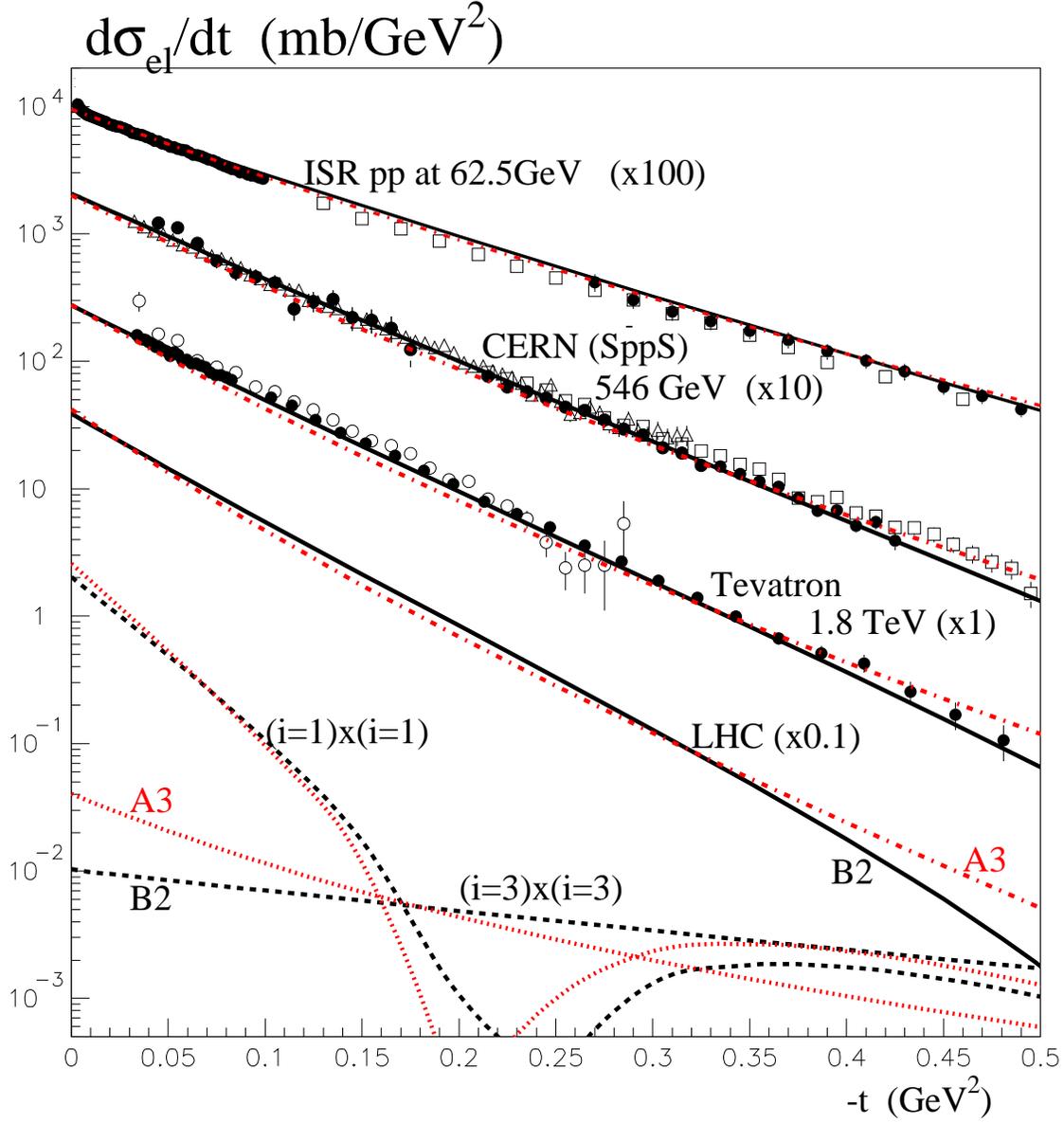}
\caption{The description of the elastic $pp$ differential cross section for models (B2) and (A3) shown by continuous and dot-dashed curves respectively. The references to the data are as in \cite{KMRsoft}.}
\label{fig:r2}
\end{center}
\end{figure}

We fit the parameter $a_1$ of (\ref{eq:V}), which specifies the form factors of the Good-Walker eigenstates $\phi_i$, and tune the value of $\alpha_\funp (0)=1+\Delta$, in order to describe the energy dependence of the measured total and elastic cross sections, and $d\sigma_{\rm el}/dt$ in the CERN-ISR to Tevatron energy range. Recall that the parameters $\gamma^2$ and $\lambda$ are specified by the available single diffractive data. A sample of the results are presented in Fig. \ref{fig:r1}\footnote{We thank Asher Gotsman for pointing out an incorrect labelling of the SD and DD curves in our previous version of the paper.} and Fig. \ref{fig:r2}.

\begin{table}[htb]
\begin{center}
\begin{tabular}{|c|c|c|c|c|c|}\hline

model  &  $\Delta$  &  $\lambda$  & $a_1$ & $\gamma^2$ & $\sigma_0$ mb   \\ \hline

(A3)  &   0.53  &  0.22  &      & 0.9  & 85 \\
(A2)  &   0.40  &  0.30  &      & 0.42 & 47 \\  
(B3)  &   0.65  &  0.30  & 1.80 & 0.48 & 38 \\
(B2)  &   0.55  &  0.33  & 1.55 & 0.275 & 33 \\  \hline
\end{tabular}
\end{center}
\caption{The values of the parameters in the various models; where $\sigma_0$ is the square of the average value of the couplings $\beta_i(0)$: that is $\sigma_0=\langle \beta_i(0)\rangle^2$. By model (A2), for example, we mean that we use version (A) of the $\phi_i$-Pomeron coupling, (\ref{eq:MA}), and we take $\sigma_{\rm SD}^{{\rm low}M}$=2 mb, see (\ref{eq:2mb}). The reason for the missing entries for $a_1$ for model (A) is explained in the text.}
\end{table}
Even without any additional tuning of the parameters, the difference between the results of the two- and three-channel eikonal models is rather small (less than about $5-10\%$), once the dispersion $\gamma^2$ of the couplings $\beta_i(0)$ is fixed. So, below, we present results just for the three-channel case. The values of the parameters are listed in Table 1.  Note that the predictions at the LHC energy are practically independent of the specific values of the vertices $\beta_i(0)$ and
of the chosen form of the $t$ dependence of $\beta_i(t)$, {\it once} we fix the value of low-mass dissociation,
$\sigma^{{\rm low} M}_{\rm SD}$ (that is the dispersion $\gamma^2$), and we satisfactorily describe the present data on the elastic cross section $d\sigma_{\rm el}/dt$. Actually the
$t$-behaviour of the vertex $V(t)$ is strongly constrained by the
data. If we choose a form different to (\ref{eq:V}), then after tuning the parameters to describe the data, the behaviour of $V(t)$ turns out to be similar to that coming from (\ref{eq:V}). Also note that since the cross section for low-mass dissociation is screened by large rescattering effects, the value of $\gamma^2$ needed to describe the data is larger than the measured ratio $\sigma_{\rm SD}^{{\rm low}M}/2\sigma_{\rm el}$, see (\ref{eq:fac2}). In particular, we require a larger $\gamma^2$ in models (A) than in models (B) to reproduce the same ratio $\sigma_{\rm SD}^{{\rm low}M}/2\sigma_{\rm el}$.

The next observation is that it proved difficult to describe the $d\sigma_{\rm el}/dt$ data using model (A), which has diffractive eigenstates of the same size. In fact we did not succeed with ansatz (\ref{eq:V}). More complicated forms were required. For model (A3) we took
\be
V(t)~=~\frac{e^{0.1t}}{(1-t/0.58)}~\frac{4m^2_N-2.9t}{4m^2_N-t},
\ee
while for model (A2) we have
\be
V(t)~=~\frac{e^{0.05t}}{(1-t/0.61)},
\ee
where $t$ is in units of $\GeV^2$. Even then, though the descriptions are reasonable, they are still not as good as those of model (B), see Fig. \ref{fig:r2}.

In summary, the description of the total cross section in Fig. \ref{fig:r1} is satisfactory, except at the lowest energies considered, where we have to add the secondary Regge contributions, which are outside our analysis. The differential elastic cross sections, shown in Fig. \ref{fig:r2}, are also well described. At the bottom of the plot we show the cross sections for the elastic scattering of two individual eigenstates $\phi_1$ and $\phi_3$ at the LHC energy. For $\phi_1$, which has a large cross section and large transverse size, we see that the diffractive dip occurs at a rather small $t$, $-t \sim 0.2~\GeV^2$. For the component $\phi_3$, with the smallest cross section, there is no dip for $-t<1 ~\GeV^2$. However, after we take the sum over all the $\phi_i \phi_k$ contributions, we observe that the resulting amplitude is structureless. Note that the interference is important. After the dip the $\phi_1 \phi_1$ amplitude changes sign. Indeed, in model (B2), for $-t>0.5 ~\GeV^2$ the whole $pp$ $d\sigma_{\rm el}/dt$ becomes smaller than that due to the small size component $\phi_3$ alone.

\begin{figure} [th]
\begin{center}
\includegraphics[height=10cm]{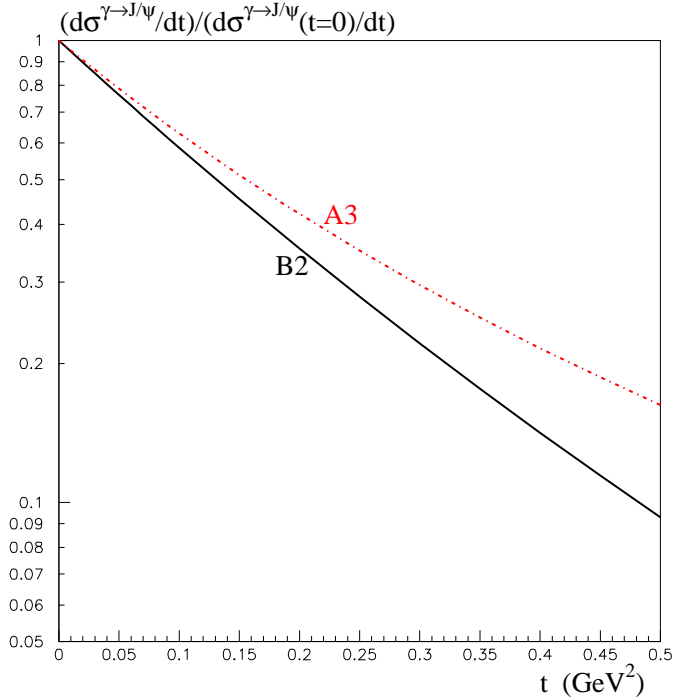}
\caption{The $t$ distribution predicted for diffractive $J/\psi$ electroproduction using models (B2) and (A3).}
\label{fig:sddis}
\end{center}
\end{figure}

The best way to check the transverse structure of the proton is to consider diffractive deep inelastic scattering (DDIS), in particular the $t$-dependence of elastic $J/\psi$ electroproduction on the proton. Here the beam particle corresponds to a `heavy photon', which acts as a small size probe of the target. In our approach, we have simply to use a beam particle eigenstate with a Good-Walker wave function of small size and small absorptive cross section. We then obtain the $t$ distribution shown in Fig. \ref{fig:sddis}. If $Q^2$ is sufficiently large, the transverse size of the incoming beam eigenfunction is much less than that of the target, so the $t$ distribution is independent of $Q^2$. Even though in model (B) we have components of quite different transverse size, for example in model (B2)
\be
R_1^2~:~R_2^2~:~R_3^2~=~1.37~:~1~:~0.26,
\ee 
the resulting $t$ distribution is very smooth. It is of the form
\be
\frac{d\sigma_{\rm J/\psi}}{dt}~ \sim~e^{Bt}~~~~~~~~{\rm with}~B=4-5~\GeV^{-2},
\ee
in agreement with H1 data \cite{h1ddis}.

\begin{figure}
\begin{center}
\includegraphics[height=17cm]{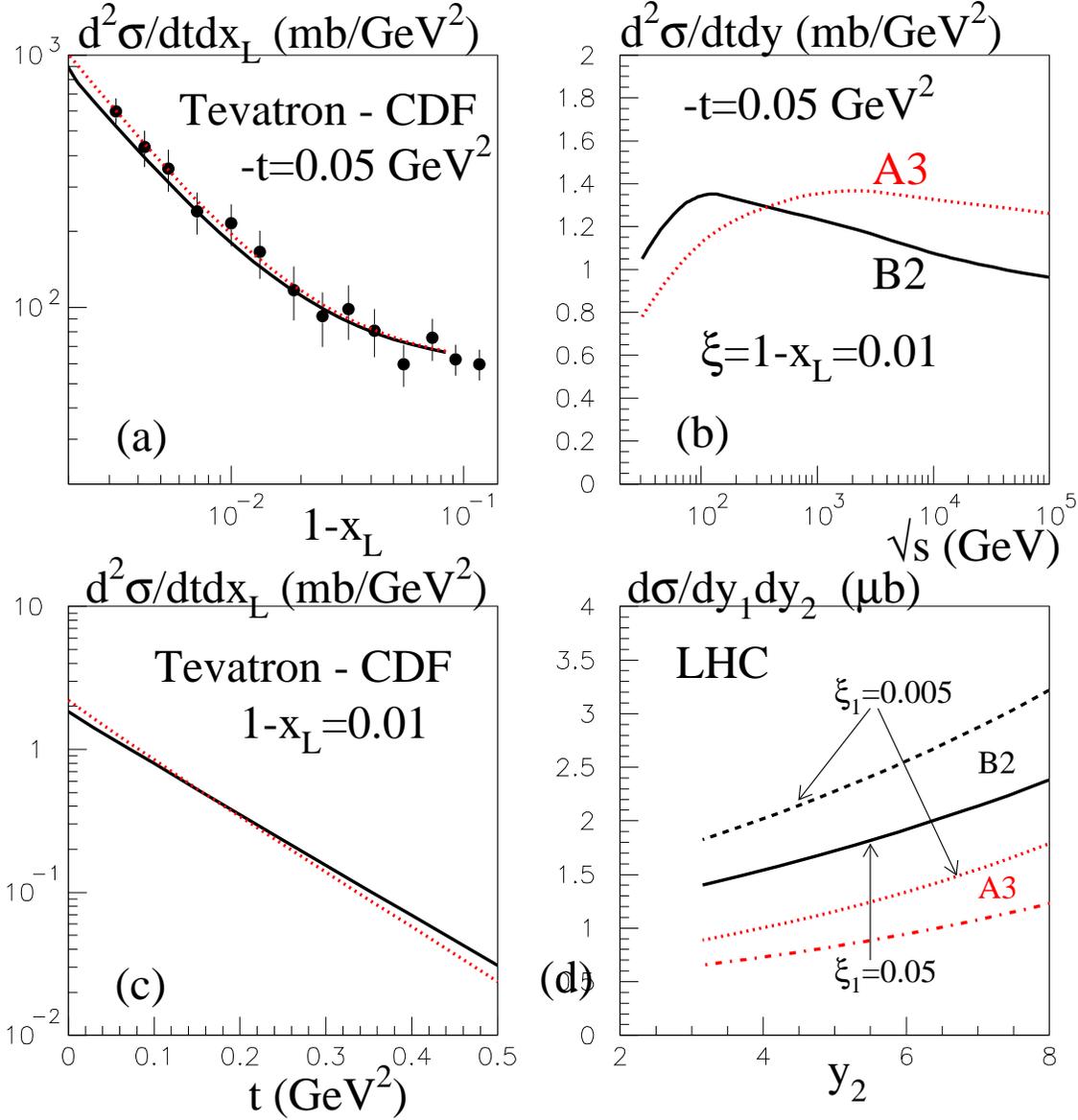}
\caption{The behaviour of the cross sections for diffractive (a,b,c) single dissociation and (d) central production obtained for two of the models, namely (A3) and (B2). The four plots are (a) the description of the CDF data \cite{CDF,CDFsd} for single dissociation; (b) the energy dependence of $d\sigma_{\rm SD}/dtdy$, where $y=-{\rm ln}\xi$ with $\xi=(1-x_L)$; (c) the $t$-dependence of $d\sigma_{\rm SD}/dtdx_L$; and (d) the $y_2$ dependence of $d\sigma_{\rm DPE}/dy_1dy_2$ for $\xi_1=0.05$ and 0.005, corresponding, respectively, to proton taggers at 220 m and 420 m from the interaction point in the LHC experiments.}
\label{fig:dsy2}
\end{center}
\end{figure}
In Fig. \ref{fig:dsy2}(a) we present the cross section for high-mass single diffractive dissociation in the kinematical region measured by CDF \cite{CDF,CDFsd}. To describe the data at relatively large $\xi \equiv (1-x_L)$ we add an $x_L$-independent secondary Regge contribution, $RR\funp$
\be
\frac{d \sigma}{dx_Ldt}~=~\frac{d \sigma^{\funp\funp\funp}}{dx_Ldt}~+~50~{\rm mb}/\GeV^2.
\ee
The last term was normalized to the large $\xi$ data and is almost negligible for $\xi<0.01$. The energy dependence for fixed $\xi=0.01$ is shown in Fig. \ref{fig:dsy2}(b). In Fig. \ref{fig:dsy2}(c) we show the $t$-dependence of this cross section. Again, despite the fact that model (B) has components of quite different radius, the $t$-dependence is well described by a single exponential with slope $B \simeq 8~\GeV^{-2}$. This is compatible with the CDF measurement \cite{CDFsd}. The major difference between the predictions of models (A) and (B) is expected in the central diffractive cross section, $d\sigma_{\rm DPE}$. This is illustrated in Fig. \ref{fig:dsy2}(d) for two values of the forward proton momentum fraction $x_L=1-\xi$, which correspond to the acceptance of roman pots placed at 420 m ($\xi=0.005$) and at 220 m ($\xi=0.05$) from the interaction point in an LHC experiment \cite{FP420,CMS-Totem}. The predicted values of the cross section have been obtained by integrating over the transverse momenta of both forward protons.

\begin{figure}
\begin{center}
\includegraphics[height=17cm]{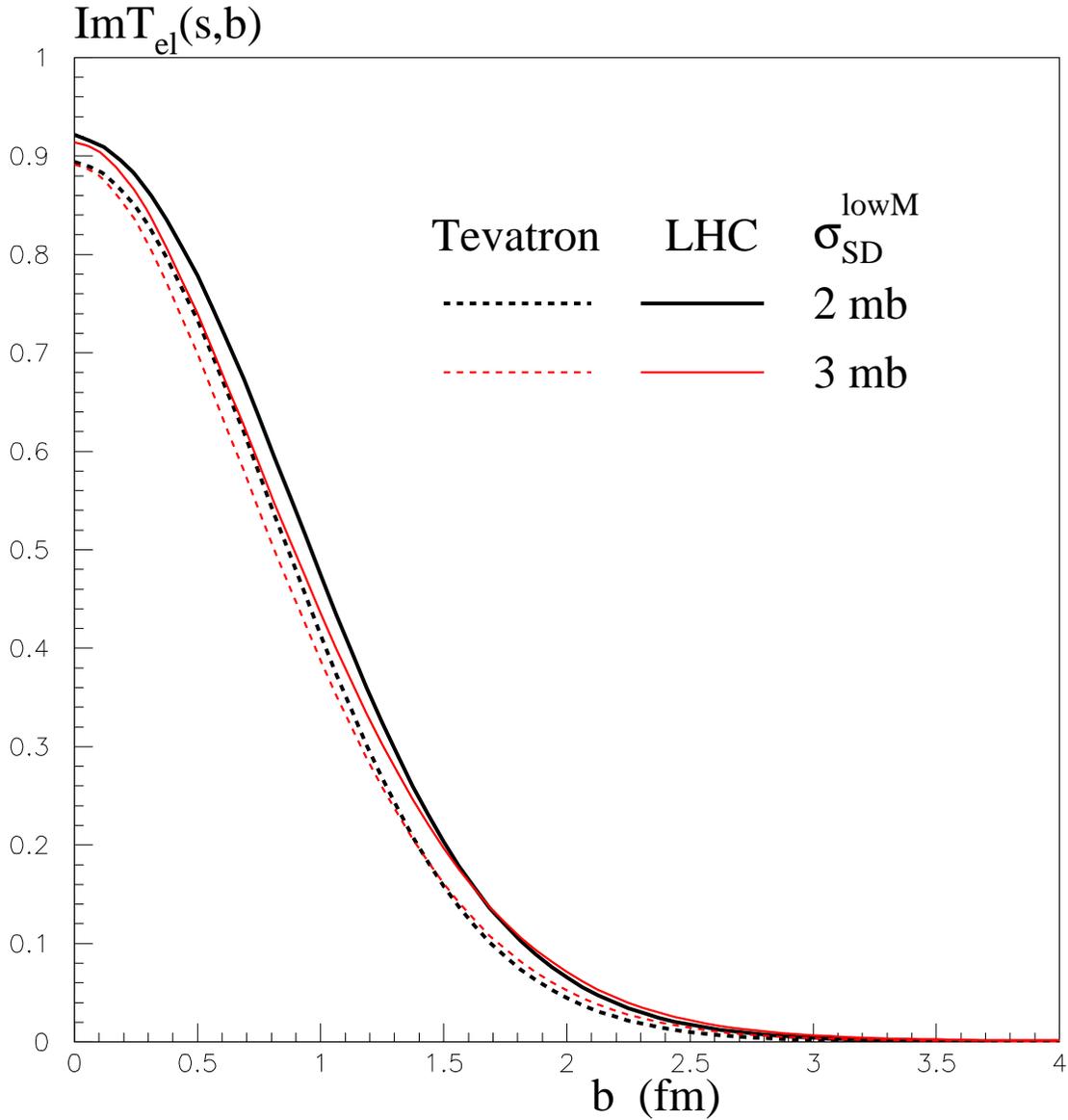}
\caption{The impact parameter shape of the elastic amplitude obtained from model (B) at the Tevatron and LHC energies. Two values, 2 and 3 mb, of $\sigma_{\rm SD}^{{\rm low}M}$ at $s \simeq 1000~\GeV^2$, have been used to fix the parameters of the model.}
\label{fig:A}
\end{center}
\end{figure}
The impact parameter shape of the elastic amplitude, $T_{\rm el}$, is shown in Fig. \ref{fig:A} for models (B2) and (B3). Since both models are tuned to describe the $d\sigma_{\rm el}/dt$ data of Fig. \ref{fig:r2} it is not surprising that the resulting amplitudes are close to each other. In fact, Im$T_{\rm el}(b)$ should be regarded as an experimentally measured quantity, up to a small correction caused by the real part of the amplitude. Indeed, we have \cite{Amaldi}
\be
{\rm Im}T_{\rm el}(b)~=~\int \sqrt{\frac{d\sigma_{\rm el}}{dt}~\frac{16\pi}{1+\rho^2}}~J_0(qb)~\frac{qdq}{4\pi},
\ee
where $q^2=|t|$ and $\rho \equiv {\rm Re}T_{el}/{\rm Im}T_{el}$. From this point of view the latest GLM model is completely inconsistent with the data, as shown by the elastic amplitudes of Fig. 10 of the first reference in \cite{GLMnew}.

Note that the energy dependence of $T_{\rm el}$ is very weak in going from Tevatron to LHC energies. This is a common feature of models which have a larger number of diffractive eigenchannels and which include high-mass diffractive dissociation. First, if we have many different eigenstates, the part of the cross section which comes from the eigenstate with the smallest cross section reaches saturation much more slowly than the other components\footnote{This observation was also noted in Refs.~\cite{GLMnew}.}.
Next, once we include the triple- (and multi-) Pomeron vertices we change the approach to saturation from the eikonal form,
\be
T~=~1-e^{-\Omega /2},
\ee
to the power-like behaviour of the type
\be
T~=~\frac{\epsilon\Omega}{1+\epsilon\Omega},
\ee
see, for example, the Schwimmer formula given in (\ref{eq:limtot}). Here we have $\Omega=\beta^2 e^{\Delta Y}=\beta^2s^\Delta$.

\begin{figure}
\begin{center}
\includegraphics[height=16cm]{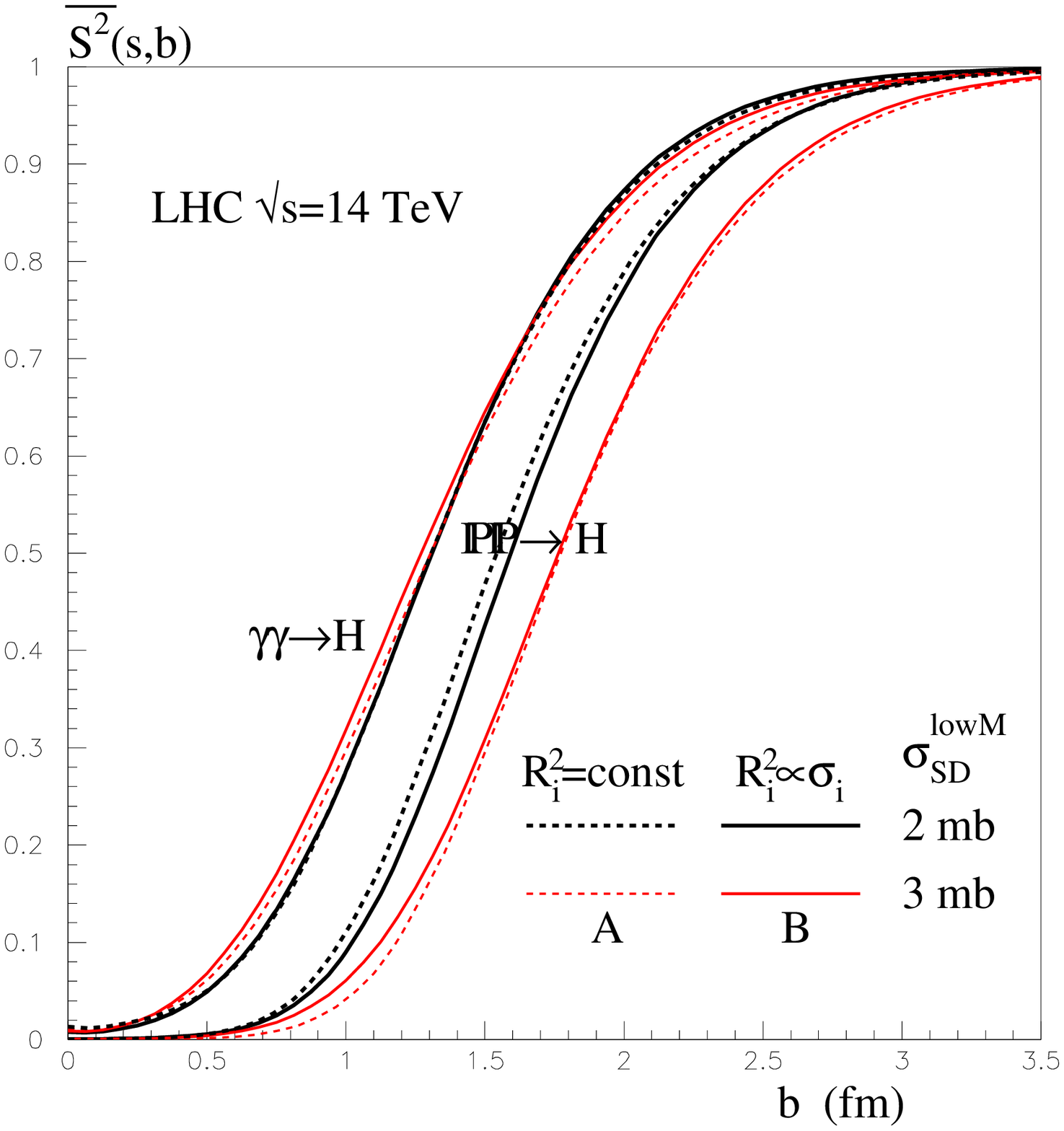}
\caption{The impact parameter behaviour of the rapidity gap survival factor $S^2$ averaged over the diffractive eigenstates $\phi_i$. Two assumptions are made about the matrix element for producing the system from eigenstates $i,k$, that is the amplitude ${\mathcal M}_{ik}$ in (\ref{eq:c3pp}). First, ${\mathcal M}_{ik} \equiv {\mathcal M}$, for which Higgs production via $\gamma\gamma$ fusion is an example. Second, ${\mathcal M}_{ik} \propto \beta_i \beta_k$, for which Pomeron-Pomeron fusion is an example. In each case the results are shown for four versions of the model: (A2), (A3), (B2) and (B3). Here the value shown for $\overline{S^2}(s,b)$ is equivalent to
$\exp(-\Omega(s,b))$ in the case of a one-channel eikonal. It
is the gap survival probability in an interaction with fixed
proton-proton impact parameter $b$. To get the complete
survival factor $\overline{S^2}$, as given in either (\ref{eq:c3}) or (\ref{eq:c3pp}), we must also average $\overline{S^2}(s,b)$ over $b$ with a weight given by the $b$-dependence of
the matrix element ${\mathcal M}(b)$.
For example, if, as in (\ref{eq:bB}), ${\mathcal M}(b)\propto \exp(-b^2/4B)$
then the full
$\overline{S^2}=\int d^2b~\overline{S^2}(s,b)\exp(-b^2/2B)/2\pi B$.}
\label{fig:A1}
\end{center}
\end{figure}
As we described in Section 3.3, the rapidity gap survival factor caused by multichannel eikonal rescattering depends on the structure of the matrix element ${\mathcal M}_{ik}$. In  Fig. \ref{fig:A1} we present the impact behaviour of the survival factor averaged over the diffractive eigenstates for two different assumptions about ${\mathcal M}_{ik}$. First we assume that ${\mathcal M}_{ik}$ does not depend on the particular eigenstate, that is ${\mathcal M}_{ik} \equiv {\mathcal M}$. An example is the exclusive production of a Higgs boson via $\gamma\gamma$ fusion. The probability to emit a photon is given just by the electric charge of the incoming state, and therefore is the same for each $\phi_i$. The corresponding curves are denoted by $\gamma\gamma \to$H on Fig. \ref{fig:A1}. The predicted $\overline{S^2(b)}$ is approximately model independent. Another possibility is to assume that ${\mathcal M}_{ik} \propto \beta_i \beta_k$, which corresponds to Higgs production by Pomeron-Pomeron fusion. The corresponding curves are denoted $\funp\funp \to$H in this case. Here we have a smaller $\overline{S^2(b)}$ since the largest contribution comes from the diffractive component with the strongest absorption. Evidently the models with the larger $\sigma_{\rm SD}^{{\rm low}M}$ give a smaller  $\overline{S^2(b)}$ due to the larger contribution of the excited ($N^*$) states to the absorptive correction.

\section{Conclusions}

We present a `new generation' model for soft high energy hadron interactions, which includes a whole series of multi-Pomeron vertices. Of course the vertices corresponding to the interactions of many Pomerons are not known. Therefore such a model may contain an infinite number of parameters. To make progress we have assumed an eikonal-like form
\be
g^n_m~ \propto ~\lambda^{n+m-2}
\ee
for the $m \to n$ Pomeron coupling. We emphasize that this assumption is much more reasonable than the assumption $g^n_m=0$ for $n+m>3$, that is keeping only the triple-Pomeron coupling $g_{3\funp}$. In fact, if we were to neglect all the higher Pomeron vertices and keep only $g_{3\funp}$, then we would obtain much stronger absorption. Indeed, the curve denoted `$g_{3\funp}$ only' in Fig. \ref{fig:r1} was calculated with exactly the same values of the parameters as in the full model (B2), and demonstrates almost complete saturation starting already at the Tevatron energy. This illustrates the important role of the higher multi-Pomeron vertices.

Another argument in favour of our approach is that it has a `transparent' partonic interpretation. All the multi-Pomeron interactions are collected in the exponential factors $e^{-(\Omega_i+\Omega_k)/2}$ which describe the absorption of intermediate partons during the evolution of the parton cascade, see (\ref{eq:Teik1}) and (\ref{eq:Teik2}).

The main goal of this paper was to formulate a new and consistent approach to high energy soft hadronic interactions. However, it is informative to check that the approach is compatible with the present data. We found that already with a minimum number of parameters (see Table 1), all versions of the model can be tuned to give a reasonable description of the available data on soft hadronic interactions in the CERN ISR-Tevatron energy range, and to give similar predictions for the LHC energy. The reason is that the energy behaviour of the amplitudes is very smooth; there are no thresholds and everything depends on ln$s$, which does not change much in going from the Tevatron to the LHC energy. Nevertheless, an important prediction of the model is the relatively low value of the total cross section expected at the LHC: $\sigma_{\rm tot} \simeq 90$ mb. This is an inescapable consequence of the absorptive corrections caused both by the low-mass intermediate states in eikonal rescattering and the high-mass excitations described by multi-Pomeron interactions\footnote{A similar value of $\sigma_{\rm tot}$ was found in a simple two-parameter phenomenological analysis \cite{sapeta}. However high-mass diffraction is not embodied in the parametric form and much less diffractive dissociation was predicted.}. The prediction for $\sigma_{\rm tot}$ is still compatible with the cosmic ray values \cite{cosmic}, although it lies below the central cosmic ray expectations.

\begin{figure}
\begin{center}
\includegraphics[height=17cm]{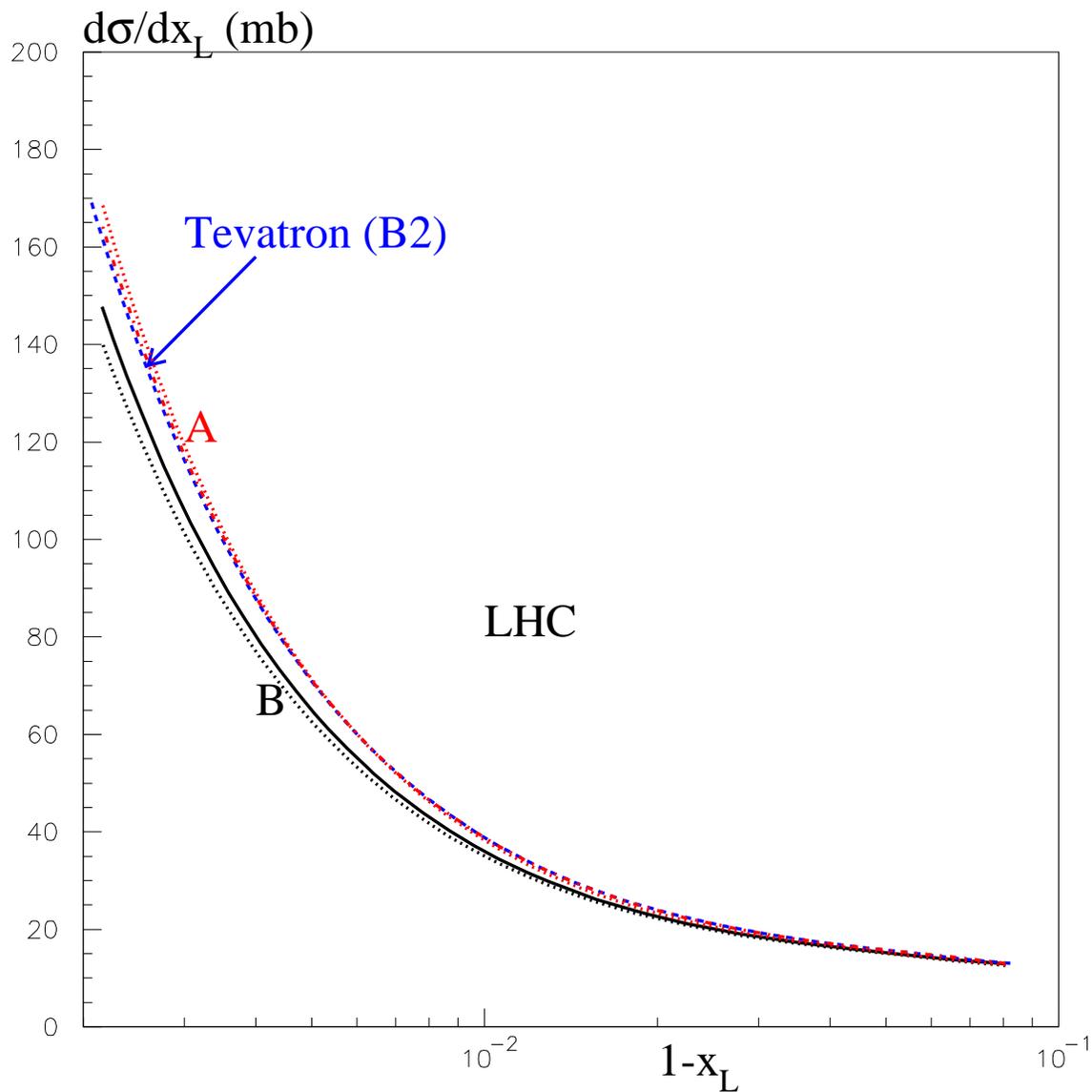}
\caption{The cross section $d\sigma_{\rm SD}/dx_L$ for single dissociation integrated over $t$ at the LHC energy resulting from four models: the continuous (dotted) curves are due to the B2 (B3) models, while the upper (red) dotted and dot-dashed curves are for models (A3) and (A2) respectively. For comparison we also show by a dashed (blue) curve the cross section obtained from model (B2) at the Tevatron. The secondary Regge contribution is included in the same way as in Fig. \ref{fig:dsy2}(a); it is relatively very small for $(1-x_L)<10^{-2}$.}
\label{fig:A2}
\end{center}
\end{figure}
Another example of very flat energy behaviour is that shown for $d\sigma_{\rm SD}/dx_L$  in Fig. \ref{fig:A2}. We see that this single dissociation cross section changes very little\footnote{The growth of the cross section caused by the Pomeron interactions $(\Delta>0)$ is compensated by stronger absorption, that is by the decrease of the gap survival factor $S^2$.} in going from the Tevatron (dashed curve) to the LHC (solid curve for model (B2)). This quantity, which is interesting in its own right, is of practical importance since it governs the rate of the so-called pile-up background to central exclusive production processes which are caused by the overlap of two soft diffractive events simultaneously with a hard scale event \cite{HKRSTW,cox}.

Note also that the rapidity gap survival factor also decreases {\it slowly} with energy. For example, we can see from the final entries of Table 2 that
\be
{\overline {S^2}}({\rm Tevatron})~/~{\overline {S^2}}({\rm LHC})~ \simeq ~1.5.
\ee
This means that the present experiments at the Tevatron can be used to check the theoretical expectations of the models at the LHC energy.

Actually the survival factor, which governs the rate of events in a particular diffractive process, must be calculated by averaging the value of $S^2(b)$ with the forms of the matrix elements ${\mathcal M}_{ik}(b)$, as in (\ref{eq:c3pp}). The values presented in Table 2 assume, for simplicity, that\footnote{Note that Ref. \cite{KMRsoft} denotes the slope as $2b=B$, where here $b$ is not to be confused with the impact parameter.}
\be
{\mathcal M}_{ik}~\propto~{\rm exp}(-b^2/4B).
\label{eq:bB}
\ee
(Of course, in general, each amplitude ${\mathcal M}_{ik}$ may have its own slope $B_{ik}$.) As examples, we choose the values $B=4$ and $5.5~\GeV^{-2}$ to be the same as in \cite{KMRsoft}. After we tune the model to the soft data, we find that we still obtain almost the same survival factors at the LHC energy; compare the entries in Table 2 with those in the ``CD(FPS)'' column of Table 1 of \cite{KMRsoft}.

\begin{table}[htb]
\begin{center}
\begin{tabular}{|l|c|c|c|c|c|}\hline

  &  Tevatron &  LHC  & $\sqrt{s}=10^5$ GeV     \\ \hline

$\sigma_{\rm tot}$ &   74.0 (73.9) & 88.0 (86.3) & 98.0 (94.3)    \\
$\sigma_{\rm el}$  &   16.3 (15.1) & 20.1 (18.1) & 22.9 (20.0)    \\ \hline

$\sigma_{\rm SD}$  &   10.9 (12.7) & 13.3 (16.1) & 15.7 (17.7)    \\
~~~$\sigma_{\rm SD}^{{\rm low}M}$  &   4.3 (6.0) & 5.1 (7.0) & 5.7 (7.9)    \\
~~~$\sigma_{\rm SD}^{{\rm high}M}$  &   6.5 (6.7) & 8.1 (9.1) & 10.0 (9.8)    \\ \hline
$\sigma_{\rm DD}$  &    7.2 (8.7)  & 13.4 (12.9) & 17.3 (21.1)    \\  
~~~$\sigma_{\rm DD}^{{\rm low}M}$  &   0.2 (0.5) & 0.2 (0.5) & 0.2 (0.6)    \\
~~~$\sigma_{\rm DD}^{{\rm high}M}$  &   4.5 (4.0) & 9.3 (5.9) & 11.7 (12.9)    \\ 
~~~$\sigma_{\rm DD}^{({\rm high}M*{\rm low}M)}$  &   2.1 (3.6) & 2.9 (5.2) & 3.8 (6.0)    \\ 
~~~$\sigma_{\rm DD}^{({\rm SD*SD})}$  &   0.4 (0.7) & 1.0 (1.3) & 1.6 (1.6)    \\ \hline
${\overline {S^2}}~~~(B=4)$  & 0.027 (0.018) & 0.017 (0.012) & 0.013 (0.009) \\ 
${\overline {S^2}}~~~(B=5.5)$  & 0.048 (0.032) & 0.032 (0.023) & 0.025 (0.018) \\ \hline
\end{tabular}
\end{center}
\caption{The cross sections (in mb) predicted at the three energies using model (B2). Also shown in brackets are the predictions for model (B3).  The various components of the single and double dissociation cross sections are also listed. The final entries are the survival factors of the rapidity gaps for exclusive diffractive central production, $pp \to p+X+p$, for two values of the slope $B=4$ and $5.5~\GeV^{-2}$. The expression for $\sigma_{\rm DD}^{({\rm SD*SD})}$ is given in (\ref{eq:DD2}).}
\end{table}
A sample of the results for the total and the diffractive dissociation cross sections were presented in Fig. \ref{fig:r1}, and for $d\sigma_{\rm el}/dt$ in Fig. \ref{fig:r2}. In Table 2 we also list the values of the cross sections for models (B2) and (B3) at three energies, namely  $\sqrt{s}=1.8~\GeV,~14$ TeV and $~10^5$ GeV. Note that model (B) is favoured, for reasons explained below. We see that with increasing energy the cross sections violate the Pumplin bound \cite{Pump}
\be
\sigma_{\rm el}~+~\sigma_{\rm SD}~+~\sigma_{\rm DD}~<~\sigma_{\rm tot}/2.
\ee
However this bound is only justified for low-mass dissociation described in terms of Good-Walker diffractive eigenstates. (Actually the bound was originally proved for dissociation on nuclei.) For the low-mass components of the diffractive dissociation cross sections we see that there is no contradiction with the bound.

An important ingredient of the analyses of soft high energy hadron interactions are the data on {\it diffractive dissociation}. The data that are available at present are fragmentary. No experiment has covered the whole kinematic range of $t$ and rapidity. Moreover, at low $M$ and low $t$ it is hard to avoid contamination from elastic events. For these reasons we prefer to compare the theoretical predictions with measurements of differential cross sections, $d\sigma_{\rm SD}^{{\rm low}M}/dt$ and $d^2\sigma_{\rm SD}^{{\rm high}M}/dtdM^2$, in regions where they were actually observed, rather than with integrated cross sections, $\sigma_{\rm SD}$ and $\sigma_{\rm DD}$, which were obtained by extrapolating the data into unmeasured regions using some simplified model.

In order to further constrain the parameters of the model so as to obtain more precise predictions, it is important to make accurate measurements at collider energies, of the single diffractive dissociation cross section for low masses $\sigma_{\rm SD}^{{\rm low}M}$ and of its $t$ dependence $d\sigma_{\rm SD}^{{\rm low}M}/dt$, and of central diffractive production $d\sigma_{\rm DPE}/dy_1dy_2$. The latter two measurements will provide information on the transverse sizes of the various diffractive eigenstates $\phi_i$. Such measurements can be performed at the LHC\footnote{We thank Risto Orava for discussion of this issue.}, for example, in the TOTEM experiment \cite{TOTEM}.

We have considered two different models for the $\phi_i$-Pomeron couplings: model (A) in which all the components have the same size, and model (B) in which the sizes differ, see (\ref{eq:MA}) and (\ref{eq:MB}). There are three observations which appear to favour model (B). {\it First}, model (B) gives a somewhat better detailed description of the various data for $d\sigma_{\rm el}/dt$. Indeed, recall for model (A3) we were required to choose a complicated form of the residue $V(t)$ to avoid a contradiction with the data. {\it Second}, Model (B) predicts $\sigma_{\rm DD}$ which is better agreement with CDF measurements\footnote{CDF\cite{CDFDD} have measured the cross section for double dissociation with a rapidity gap $\Delta y>3$ enclosing $y=0$ and find $\Delta\sigma_{\rm DD}=4.4 \pm 1.2$ mb at 1.8 TeV and $3.4 \pm 1.1$ mb at 630 GeV, as compared to the predictions of $\Delta\sigma_{\rm DD}=$4.4 mb and 3.7 mb, respectively, of model (B2). Here $\Delta\sigma_{\rm DD}$ denotes the portion of the cross section for double dissociation coming from the kinematic region where it was actually measured.}. {\it Third}, model (A) requires a very large dispersion $\gamma^2$ of the Pomeron-$\phi_i$ couplings, which tends to be incompatible with the data for soft reactions at fixed target energies, see \cite{KAID}. Nevertheless, at present we cannot completely reject model (A), and so we have also presented the results of this model to indicate the possible spread of the predictions.

In all the models the value of $\lambda$ was about 1/3, in agreement with the evaluation of $g_{3\funp}$ based on the HERA data for $\gamma p \to J/\psi~X$ \cite{KMRj}.
 
In our description of high energy `soft' hadron-hadron interactions we need to start with a rather large `bare Pomeron' intercept: $\alpha_\funp (0)=1+\Delta$ with $\Delta \sim 0.5$. Note that this is considerably larger than the intercept $\Delta \sim 0.08$ of the naive effective Donnachie-Landshoff-type of Pomeron \cite{DL}. The reason is that it is necessary to compensate the effects of the screening corrections originating from the enhanced multi-Pomeron diagrams. Recall that in the present model we use the heavy Pomeron approximation \cite{gribovh}, that is we set the slope of the Pomeron trajectory $\alpha'=0$. Nevertheless this extreme choice is still consistent with the data. The shrinkage of the elastic differential cross section data of Fig. \ref{fig:r2} is reproduced in our approach by a stronger absorptive effect at low impact parameters with increasing energy. Our preliminary estimates indicate that a small value of $\alpha' (\alpha' \lapproxeq 0.05 ~\GeV^{-2})$ may possibly improve the description of the data\footnote{Simultaneously, the inclusion of a non-zero $\alpha'$ will result in a smaller value of $\Delta$.}, but we find a larger $\alpha'$ ($\alpha' \gapproxeq 0.05~\GeV^{-2}$) appears to be definitely ruled out, since the main shrinkage of the elastic peak comes from absorptive effects.

\begin{figure}
\begin{center}
\includegraphics[height=4cm]{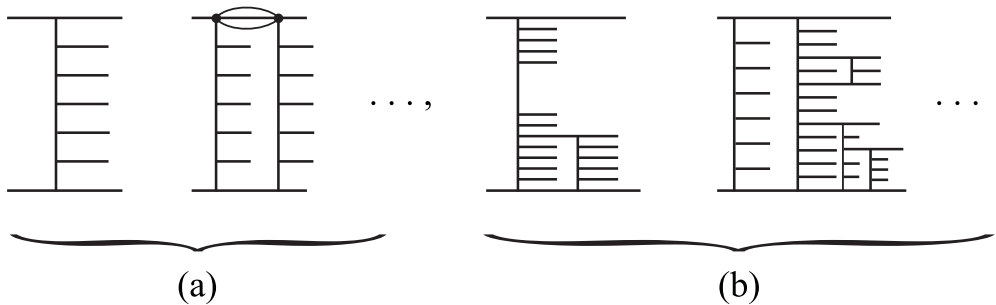}
\caption{Typical final states which are included in the model: (a) diagrams from multi-channel eikonal rescattering, (b) diagrams with and without a rapidity gap from enhanced multi-Pomeron interactions.}
\label{fig:agk}
\end{center}
\end{figure}
In summary, we have been able, for the first time, to present a theoretically fully consistent description of high energy soft hadron interactions. The procedure incorporates both a multi-channel eikonal and multi-Pomeron interactions. In terms of Reggeon Field Theory, it sums up the eikonal diagrams in  Fig. \ref{fig:fF}, together with the enhanced multi-Pomeron diagrams like those in Figs. \ref{fig:both} and \ref{fig:loops}. In terms of final states these diagrams generate processes with different densities of secondary particles. The eikonal diagrams generate processes like those shown in Fig. \ref{fig:agk}(a), while the enhanced multi-Pomeron diagrams may generate processes with rapidity gaps, as well as those with different densities of secondaries in particular rapidity intervals as shown in Fig. \ref{fig:agk}(b); see the discussion in Section 4.3.

\section*{Acknowledgements}

We thank Aliosha Kaidalov and Sergey Ostapchenko for valuable discussions. MGR thanks the IPPP at the University of Durham  for Physics for hospitality. This work was supported by INTAS grant 05-103-7515, by grant RFBR 07-02-00023, by the Russian State grant RSGSS-5788.2006.02, and by the Russia-Israel grant 06-02-72041-204; 205; 210; 200.


\begin{thebibliography}{xx}  
 
\bibitem{DL} A. Donnachie and P.V. Landshoff, Phys. Lett. {\bf B296} (1992) 227. 
   
\bibitem{Good} M.L. Good and W.D. Walker, Phys. Rev. {\bf 120}, 1857 (1960); \\
E.L. Feinberg and I.Ya. Pomeranchuk, Doklady Akad. Nauk SSSR  {\bf 93}, 439 (1953); Suppl. Nuovo Cimento v.~{\bf III}, serie~X, 652 (1956). 

\bibitem{kaid} A.B. Kaidalov, Phys. Rep. {\bf 50} (1979) 157.    

\bibitem {tmr}K.A. Ter-Martirosyan, ITEP preprints No.\ 70, 71 (1975); 7, 11, 133--135,     
158 (1976).  

\bibitem {uri}  E.~Gotsman, E.~Levin, U.~Maor, E.~Naftali and A.~Prygarin,
               hep-ph/0511060;\\ 
               published in the proceedings of the workshop
               {\it HERA and the LHC: A Workshop on the Implications of 
               HERA for LHC Physics}, 
               hep-ph/0601012, p.\ 221 and references therein.

\bibitem {bh} M.M. Block, Phys. Rept. {\bf 436} (2006) 71 and references therein.

\bibitem{KMRsoft} V.A. Khoze, A.D. Martin and M.G. Ryskin, Eur. Phys. J. {\bf C18} (2000) 167.

\bibitem {islam} M.M. Islam, R.J. Ruddy and A.V. Prokudin, Phys. Lett. {\bf B605} (2005) 115; Int. J. Mod. Phys. {\bf A21} (2006) 1.

\bibitem {GLMnew} E. Gotsman, E. Levin and U. Maor, arXiv:0708.1506 (2007) ;\\
E.~Gotsman, A.~Kormilitzin, E.~Levin and U.~Maor, Eur. Phys. J. {\bf C52} (2007) 295. 

\bibitem {KMRj} V.A. Khoze, A.D. Martin and M.G. Ryskin, Phys. Lett. {\bf B643} (2006) 93.


\bibitem {kkt} A.B. Kaidalov, V.A. Khoze, Yu.F. Pirogov and N.L. Ter-Isaakyan, Phys.     
Lett. {\bf B45} (1973) 493; \\    
A.B. Kaidalov and K.A. Ter-Martirosyan, Nucl. Phys. {\bf B75} (1974) 471.

\bibitem {ff}R.D. Field and G.C. Fox, Nucl. Phys. {\bf B80} (1974) 367.

\bibitem {abram} V.A. Abramovsky, Pis'ma v JETP {\bf 23} (1976) 228; \\
V.A. Abramovsky and R.G. Betman, Sov. J. Nucl. Phys. {\bf 49} (1989) 747; \\
V.A. Abramovsky, A.V. Dmitriev and A.A. Schneider, hep-ph/0512199. 

\bibitem {KPT} A.B. Kaidalov, L.A. Ponomarev and K.A. Ter-Martirosyan, Sov. J. Nucl.     
Phys. {\bf 44} (1986) 468.

\bibitem {schw} A. Schwimmer, Nucl. Phys. {\bf B94} (1975) 445.

\bibitem {glr}  L.~V.~Gribov, E.~M.~Levin and M.~G.~Ryskin,
%``Semihard Processes In QCD,''
Phys.\ Rept.\  {\bf 100} (1983) 1.

\bibitem {bk}I. Balitsky, Nucl. Phys. {\bf  B463} (1996) 99;
Phys. Rev. Lett. {\bf 81} (1998) 2024; Phys. Lett. {\bf B518} (2001) 235.\\
Y.V. Kovchegov, Phys. Rev. {\bf D60} (1999) 034008;
Phys. Rev. {\bf D61} (2000) 074018.

\bibitem {afs}
D. Amati, L. Caneschi and R. Jengo, Nucl.\ Phys.\ {\bf B101} (1975) 397;\\
V. Alessandrini, D. Amati and R. Jengo, Nucl.\ Phys.\ {\bf B108} (1976) 425;\\
 D.~Amati, M.~Le Bellac, G.~Marchesini and M.~Ciafaloni,
Nucl.\ Phys.\ {\bf B112} (1976) 107;\\
D.~Amati, G.~Marchesini, M.~Ciafaloni and G.~Parisi,
 %``Expanding Disk As A Dynamical Vacuum Instability In Reggeon Field Theory,''
Nucl.\ Phys.\ {\bf B114}, 483 (1976).


\bibitem {ost} S. Ostapchenko, Phys. Rev. {\bf D74} (2006) 014026;
 Phys. Lett. {\bf B636} (2006) 40; hep-ph/0612175.

\bibitem {motyka} S. Bondarenko and L. Motyka,   Phys.\ Rev.\ {\bf D75} (2007) 114015.

\bibitem {KMRcalabria} V.A.~Khoze, A.D.~Martin and M.G.~Ryskin,
  %``Soft diffraction at the LHC and properties of the pomeron,''
  Nucl.\ Phys.\ Proc.\ Suppl.\  {\bf 99B} (2001) 213.

\bibitem {KMRprosp} V.A. Khoze, A.D. Martin and M.G. Ryskin, Eur. Phys. J. {\bf C14} (2000) 525; 
Eur. Phys. J. {\bf C23} (2002) 311.

\bibitem{HKRSTW} S. Heinemeyer {\em et al.}, arXiv:0708.3052 [hep-ph] (2007).

\bibitem{KKMRprob} 
A.B. Kaidalov, V.A. Khoze, A.D. Martin and M.G. Ryskin, Eur. Phys. J. {\bf C21} (2001) 521. 


\bibitem {glm} S. Bondarenko, E. Gotsman, E. Levin and U. Maor,
Nucl. Phys. {\bf A683} (2001) 644.

\bibitem {BKKMR}  K.G.~Boreskov, A.B.~Kaidalov, V.A.~Khoze, A.D.~Martin and M.G.~Ryskin,
Eur.\ Phys.\ J.\ {\bf C44} (2005) 523;\\
A.B. Kaidalov, hep-ph/0612358.

\bibitem {CDF}
K.~Goulianos and J.~Montanha,
  %``Factorization and scaling in hadronic diffraction,''
  Phys.\ Rev.\ {\bf D59} (1999) 114017.

\bibitem{CDFsd}  F.~Abe {\it et al.}  [CDF Collaboration],
%``Measurement of $\bar{p}p$ single diffraction dissociation at $\sqrt{s} =
%546$ GeV and 1800 GeV,''
Phys.\ Rev.\ {\bf D50} (1994) 5535.

\bibitem {KMRjhep} V.A. Khoze, A.D. Martin and M.G. Ryskin, JHEP {\bf 0605}:036 (2006).

\bibitem {Cardy} J.L. Cardy, Nucl. Phys. {\bf B75} (1974) 413.

\bibitem {KAT} M.S.~Dubovikov and K.A.~Ter-Martirosian,
  %``Theory Of The Froissaron Exchange,''
  Nucl.\ Phys.\ {\bf B124}, 163 (1977);\\
M.S.~Dubovikov, B.Z.~Kopeliovich, L.I.~Lapidus and K.A.~Ter-Martirosian,
  %``Dynamics Of Froissarons In High-Energy Physics,''
  Nucl.\ Phys.\ {\bf B123} (1977) 147.

\bibitem{fey} R.P. Feynman, Phys. Rev. Lett. {\bf 23} (1969) 1415.

\bibitem{Gribbook}  V.N. Gribov, in {\it Gauge Theories and Quark Confinement}, PHASIS, Moscow, 2002, p.3.

\bibitem{amati} D. Amati, A. Stanghellini and S. Fubini, Nuovo Cim. {\bf 26} (1962) 896.

\bibitem {lr} E.M. Levin and M.G. Ryskin, Nucl. Phys. {\bf B304} (1988) 805.

\bibitem{agk} V.A. Abramovsky, V.N. Gribov and O.V. Kancheli, Sov. J. Nucl. Phys. {\bf     
18} (1974) 308.

\bibitem{KLO} V.A.~Khoze, S.~Lupia and W.~Ochs, Phys.\ Lett.\  {\bf B394} (1997) 179.
 
\bibitem {gribovh} V.N.~Gribov,
%``A Theory Of The Heavy Pomeron,''
Nucl.\ Phys.\  {\bf B106} (1976) 189.

%\bibitem {BFKLrev} L.N Lipatov, Phys. Reports
%  L.~N.~Lipatov,
 %``Small-x physics in perturbative QCD,''
%Phys.\ Rept.\  {\bf 286} (1997) 131;\\
%G.~P.~Salam,
%``A resummation of large sub-leading corrections at small x,''
%JHEP {\bf 9807} (1998) 019;\\
%M.~Ciafaloni, D.~Colferai and G.P.~Salam,
%Phys.\ Rev.\ {\bf D60}, 114036 (1999);\\
%G.P.~Salam,
%``An introduction to leading and next-to-leading BFKL,''
%Acta Phys.\ Polon.\  B {\bf 30}, 3679 (1999);\\
%V.A.~Khoze, A.D.~Martin, M.G.~Ryskin and W.J.~Stirling,
  %``The spread of the gluon k(t)-distribution and the determination of the
  %saturation scale at hadron colliders in resummed NLL BFKL,''
%  Phys.\ Rev.\ {\bf D70} (2004) 074013.

\bibitem {KMRenhanced}
B. Andersson, G. Gustafson and J. Samuelsson, Nucl. Phys. {\bf B467} (1996) 443;\\
J. Kwieci\'{n}ski, A.D. Martin and P.J. Sutton, Z. Phys. {\bf C71} (1996) 585;\\
M. Ciafaloni, D. Colferai and G. Salam, Phys. Rev. {\bf D60} (1999) 114036; \\ 
C.R. Schmidt, Phys. Rev. {\bf D60} (1999) 074003;\\
J.R.~Forshaw, D.A.~Ross and A.~Sabio Vera, Phys. Lett. {\bf B455} (1999) 273;\\
G. Chachamis, M. Lublinsky and A. Sabio Vera, Nucl. Phys. {\bf A748} (2005) 649.

\bibitem{KAID} A.B. Kaidalov, Yad. Fiz.{\bf 13} (1971) 401 (in Russian)    

\bibitem{reson} L.~Baksay {\it et al.}, Phys.\ Lett.\ {\bf B53} (1975) 484; \\
R.~Webb {\it et al.}, Phys.\ Lett.\ {\bf B55} (1975) 331; \\
L.~Baksay {\it et al.}, Phys.\ Lett.\ {\bf B61} (1976) 405; \\
H.~de Kerret {\it et al.}, Phys.\ Lett.\ {\bf B63} (1976) 477; \\
G.C.~Mantovani {\it et al.}, Phys.\ Lett.\ {\bf B64} (1976) 471.

\bibitem {ISR} 
J.C.M.~Armitage {\it et al.},
 %``Diffraction dissociation in proton proton collisions at {ISR} energies,''
 Nucl.\ Phys.\ {\bf B194} (1982) 365.

\bibitem{Albrow} M. Albrow {\em et al.}, Nucl. Phys. {\bf B108} (1976) 1; \\
Y. Akimov {\it et al.}, Phys. Rev. Lett. {\bf 35} (1975) 763, 766.

\bibitem{h1ddis} H1 Collaboration, A. Aktas {\em et al.},
  Eur. Phys. J. {\bf C46} (2006) 585; \\
ZEUS Collaboration, S. Chekanov {\em et al.},
  Eur. Phys. J. {\bf C24} (2002) 345; \\
ZEUS Collaboration, S. Chekanov {\em et al.},
  Nucl. Phys. {\bf B695} (2004) 3. 


%\bibitem{BLOCK} See, M.M. Block, K. Kang and A.R. White, Int. J. Mod. Phys. {\bf A7}     
%(1992) 4449, for a collection and discussion of elastic data.    
%\bibitem{AG} A.A. Anselm and V.N. Gribov, Phys. Lett. {\bf B40} (1972) 487.    
%\bibitem{COLLINS} P.D.B. Collins, Regge Theory and High Energy Physics, Cambridge     
%Univ. Press (1977).    
%\bibitem{AGK} V.A. Abramovsky, V.N. Gribov and O.V. Kancheli, Sov. J. Nucl. Phys. {\bf     
%18} (1974) 308.    
%\bibitem {KKMRln} A.B. Kaidalov, V.A. Khoze, A.D. Martin and M.G. Ryskin, Eur. Phys. J. {\bf C47} (2006) 385.
%\bibitem{abramovsky} V.A. Abramovsky

%\bibitem{KG} K. Goulianos, Phys. Rev. {\bf D14} (1976) 1445.    
%\bibitem{G} V.N. Gribov, Sov. J. Nucl. Phys. {\bf 17} (1973) 313.    

%\bibitem{KMRtag} V.A. Khoze, A.D. Martin and M.G. Ryskin, Eur. Phys. J. {\bf C24} (2002) 581.
\bibitem{FP420} M.~Albrow et al.,
                CERN-LHCC-2005-025, June 2005.

\bibitem{CMS-Totem} CMS and TOTEM diffractive and forward physics
                    working group,
                    CERN/LHCC 2006-039/G-124,
                    CMS Note 2007/002, TOTEM Note 06-5,
                    December 2006.

\bibitem{Amaldi} U.Amaldi, {\it  Elastic and inelastic processes at the Intersecting Storage Rings. The Experiments and their impact parameter description} in Erice 1973, Proceedings, Laws of Hadronic Matter, New York 1975, 673-741.

\bibitem{sapeta} S. Sapeta and K.J. Golec-Biernat, Phys. Lett. {\bf B613} (2005) 154. 

\bibitem{cosmic} COMPAS Group, IHEP, Protvino, Russia, August 2005; \\
J.R.~Cudell {\it et al.,}  [COMPETE Collaboration], Phys.\ Rev.\ Lett.\  {\bf 89} (2002) 201801.

\bibitem{cox}  B.E. Cox, F.K. Loebinger and A.D. Pilkington, arXiv:0709.3035 [hep-ph] (2007).

%\bibitem {h1ddis} H1 Collaboration: A. Aktas {\it et al.} Eur. Phys. J. {\bf C48} (2006) 749. 

\bibitem{Pump} J.~Pumplin, Phys.\ Rev.\ {\bf D8} (1973) 2.

\bibitem{TOTEM} V.~Berardi {\it et al.,}\ [TOTEM Collaboration],
                TDR, CERN-LHCC-2004-002, TOTEM-TDR-001, January 2004.

\bibitem{CDFDD} T. Affolder {\em et al.}, Phys. Rev. Lett. {\bf 87} (2001) 141802.


\end{thebibliography}
\end{document}